# A Novel Sharp Interface Immersed Boundary Framework for Viscous Flow Simulations at Arbitrary Mach Number Involving Complex and Moving Boundaries

Pradeep Kumar Seshadri[1], Ashoke De*[1]

[1]Department of Aerospace Engineering, Indian Institute of Technology Kanpur, 208016, Kanpur, India

## Abstract

This work presents a robust and efficient sharp interface immersed boundary (IBM) framework, which is applicable for all-speed flow regimes and is capable of handling arbitrarily complex bodies (stationary or moving). The work deploys an in-house, parallel, multi-block structured finite volume flow solver, which employs a 3D unsteady Favre averaged Navier Stokes equations in a generalized curvilinear coordinate system; while we employ a combination of HCIB (Hybrid Cartesian Immersed boundary) method and GC(Ghost-cell) for solution reconstruction near immersed boundary interface. A significant difficulty for these sharp interface approaches is of handling sharp features/edges of complex geometries. In this study, we observe that apart from the need for robust node classification strategy and higher order boundary formulations, the direction in which the reconstruction procedures are performed plays an important role in handling sharp edges. Taking this into account we present a versatile interface tracking procedure based on ray tracing algorithm and a novel three step solution reconstruction procedure that computes pseudo-normals in the regions where the normal is not well-defined and reconstructs the flow field along those directions. We demonstrate that this procedure enables solver to efficiently handle and accurately represent sharp-edged regions. A fifth-order weighted essentially non-oscillatory (WENO) scheme is used for capturing shock-induced discontinuities and complex fluid-solid interactions with high resolution. The developed IBM framework is applied to a wide range of flow phenomena encompassing all-speed regimes (M=0.001 to M = 2.0). A total of seven benchmark cases (three stationary and four moving bodies) are presented involving various geometries (cylinder, airfoil, wedge) and the predictions are found to be in excellent agreement with the published results.

*Corresponding author, E-mail: ashoke@iitk.ac.in, Ph: +91-512-6797863, Fax : +91-512-6797561

# 1. Introduction

In past decade immersed boundary methods has emerged as a well-established framework for simulating flows involving geometrically complex objects either stationary or in motion especially for incompressible flows[1-6] and high-speed flow phenomena [7-10] where the effects of the viscous boundary layer are neglected . In recent years, efforts are being made to extend the immersed boundary framework for high Reynolds number applications [11-14] as well as to compressible viscous flow problems [15-21]. Despite the exponential increase in immersed boundary related literature, only scarce resources are available on the applicability and efficacy of immersed boundary framework to the flow fields where both compressible flows and weakly compressible flows (M<0.2) exists simultaneously.

In general, most of the flow solvers are designed for handling only a specific flow regime. For example, compressible flow solvers face the loss of convergence and accuracy in the incompressible limit due to large condition number arising from the large disparity in acoustic and particle velocities. But it is not possible for compressible flow solvers to entirely avoid encountering low Mach region when dealing with say transonic flow fields or combustion related flow phenomena. A careful survey of the existing literature on immersed boundary framework for compressible flows suggests few things:

**(1)** Most of the studies involving compressible flow solvers restrict themselves to Mach number regime of 0.2 or above. The work of Khalili et al. [18] is an exception. They had simulated inline oscillating cylinder in a quiescent medium at a reference Mach number of 0.03. Muralidharan and Menon [22] reported minor pressure oscillations for moving cylinder case at Re = 40 and Mach number 0.2 despite their higher-order adaptive cut cell approach. When they simulated transversely oscillating cylinder case at Re = 185 for same

Mach number, these oscillations were not observed. More interesting is the observation of Bharadwaj et al. [23] for a transonic flow past NACA16009 airfoil at a Mach number of 0.75, where their mass conserved sharp interface formulation showed more oscillations in their pressure distribution plots, especially near the leading edge. But these oscillations are discarded as data noise. All these observations from various studies suggest that there is a lack of clarity regarding the source of these errors: whether the oscillations are due to limits of the flow solver or the inadequacy of immersed boundary framework to handle such flows.

**(2)** Traditionally for simulating low speed flows using compressible flow solvers; the general approach is to employ preconditioning techniques. A handful of research works do exist that had incorporated immersed boundary framework in their precondition based compressible flow solvers. de Tuillo and co-workers [15, 24] were the first to utilize precondition based flow solver for immersed boundary based simulations. They reported low Mach number flows past stationary objects. Lv et al. [25] utilized unstructured multi-grid preconditioned flow solver for simulating moving body problems involving unsteady compressible flow. They used an approach called immersed membrane method (IMM) which reconstructs the flow along the cell edges. Tyagi et al.[26],[27] and Roy et al.[28],[29] have used precondition based density flow solvers for simulating impeller stirred tank related problems which involved high Reynolds number turbulent regime. None of these studies report on issues of mass conservation and spurious oscillations encountered in simulating moving body problems. To the best of author's knowledge, the recent study by Al-Marouf et al. [30] on adaptive mesh based embedded ghost fluid method is the only study that has extensively reported on static and moving body problems involving all speed flow regime. They use a flow solver that is based on unsplit second order Gudonov method of Collela [31]. The lowest Mach number that is reported in all these studies is of the order $O(10^{-2})$.

The brief review cited above suggests that there is a need for rigorous analysis of immersed boundary framework for its efficacy and applicability in the handling of all speed flow regime. The main objective of this study is to develop a simple and robust sharp interface based immersed boundary framework for simulating viscous flows at arbitrary Mach number and rigorously validate stationary and moving problems against benchmark studies for a wide class of flow configurations. We make use of our extensively validated in-house preconditioned flow solver [32] that is applicable to all-speed flows for this purpose.

A major challenge in almost all variants of the immersed boundary approach is regarding how to efficiently handle slender geometries or sharp edges. The diffused interface approach [33], as the name suggests, smears out the sharp features of the complex geometries over a number of Eulerian cells. This altering of geometry at times can modify the pressure distribution around the body resulting in a different flow physics than what is intended to be captured. Kang et al. [34] reported that such smeared out pressure profiles can cause parasitic currents when they are used for making velocity field divergence free.

In case of sharp interface approach, the challenge involved is twofold: One, the infinite curvature at the sharp edges makes the traditionally used signed distance (SD) based node classification algorithm inconsistent. Several improvements in such algorithms have been proposed [35-38] and many have adopted ray tracing (RT) based node classification algorithms [30, 39]. Some even opt for special ad-hoc treatments [40]. The second challenge involves the lack of enough grid resolution to capture the sharp features. Some have reduced their order of accuracy of their interpolation schemes [38] while some resort to higher order boundary formulation to increase the order of accuracy [41, 42]. Despite such efforts many studies claim unsatisfactory results by using sharp interface approach for modelling flow past airfoils with sharp trailing edges. In a very recent work, Menon and Mittal [43] have used airfoil with slightly rounded trailing edge for modelling flow induced pitching oscillations at Reynolds number 1000. They reported that such rounding ensures that flow is well resolved around it. Hartung and Gilge [44] investigated scale resolving simulations of turbulent flow channels over riblet structures that are characterized by sharp edges and reported that body fitted grid simulations performed better than immersed boundary approach. In turbulent flows, capturing thin boundary layer over those sharp edges is very much essential for accurate modelling of flow physics. The body-fitted grid literature has well documented studies analyzing blunt and sharp edged trailing edges which report that modification of geometries at both low and high Reynolds number can significantly impact the flow phenomena [45-47]. Geng et al. [48] reported that trailing edge geometry is a crucial factor in determining the trailing edge vortex dynamics in a dynamic stall phenomenon. Modifying the geometry can alter the flow physics drastically. This considerations have led to development of hybrid approach between immersed boundary and overlapping/overset grids [49, 50] preserving the best features of both the frameworks.

Increasingly cut cell approach [17, 51-53] and adaptive mesh refinement strategy [24, 30, 54, 55] or even hybrid of both [19, 22] are being used to address the limitations discussed above. While these approaches have drastically improved the accuracy of the IB approach in high Reynolds number flows, robust handling of sharp edges remains an issue. For instance, one of the basic assumptions in cut-cell approach is that the immersed surface can cut the fluid cells only once. Sharp edges often violate this assumption as it can lead to multiple cuts. Usually under such scenarios, the multiple cuts are approximated to a single cutting plane by curve-fitting procedure. But handling topology of such cut cells effectively can be very challenging. Hartmann et al. [17] reported that in order to compute the geometry of cut-cells intersected by sharp corners they use marching cubes algorithm. They use a lookup table that identifies the correct configuration of the cut cell out of a list of 256 possible polygon configurations. Handling split cells which divide the cut cells into unconnected parts can be much more challenging. On the other hand, with adaptive mesh refinement approach, sharp edges are handled only by refining the grids locally. Despite this approach, due to Cartesian nature of these grids such sharp edges with infinite curvature get regularized beyond certain level of refinement. Thus, representing sharp edges in an immersed boundary framework is still a challenge that needs to be addressed.

The novelty of our proposed study is our new sharp-interface reconstruction algorithm that efficiently handles complex geometries with sharp edges providing a well resolved flow field and improved solution accuracy. While all of the studies reviewed above focuses on efficient node classification and higher order interpolation schemes as the two most important factors influencing the accuracy of the sharp-interface approach, through this work we show that there is a third factor that is equally if not more influential than the other two, namely the direction along which the interpolation stencils are applied to impose the boundary conditions. Our approach can be classified as the hybrid between HCIB (hybrid Cartesian immersed boundary) method introduced by Gilmanov et al. [6, 56] and GC (ghost-cell) approach introduced by Ghais et al.[16]. We apply interpolation/extrapolation schemes along the normal direction reconstructing the flow field variables at the immersed boundary (IB) nodes/ghost nodes (GN) for modelling, stationary and moving body problems. But as discussed earlier, sharp edges do not have well defined normal which is a serious limitation of these approaches while handling complex geometries. Both signed distance (SD) node classification and surface normal (SN) reconstruction fail to perform consistently under such circumstances. Even under the normal circumstances, the unstructured triangular mesh elements representing immersed bodies do not have well defined normal along their edges and their vertices as they are $C^0$ continuous elements [57]. So for any IB/GN node that is closer to edges or vertices, where it's normal is not well-defined, flow field variables are reconstructed in a direction that is parallel to the surface normal. This actually distorts the representation of the immersed body. In order to overcome this limitation, we have adopted the idea of defining angle-weighted pseudo normal [57] along edges of the triangle element and on its vertices making every region on the immersed body well oriented. Before we apply reconstruction procedure, we identify whether a given IB node or GN node lies closer to the triangle surface or edge or vertices with the help of computational geometry algorithms and then perform reconstruction in a direction parallel to local normal (LN), i.e. it can be a face, edge or vertex normal whichever is closer to those nodes. Many have utilized the idea of pseudo normals [35-37], but only for the purpose of improving node classification algorithms. None have applied it to determine the reconstruction direction. Through this study, we show that this approach helps in accurately modeling the sharp edges.

A recent study by Haji Mohammadi et al. [58] appreciates that the reconstruction direction is an important factor in determining solution accuracy and points out that applying reconstruction along a specific direction, i.e. either along grid lines [35, 59-61] or local surface normal direction [3, 6, 16, 35, 56] for incompressible Navier-Stokes equations with strong elliptic characteristics will result in loss of solution accuracy. They have used a moving least squares framework (which considers the information available in all the spatial directions) for reconstructing the flow field near the immersed surface satisfying the boundary conditions accurately. Similar reasoning applied to our context of all-speed flow solver suggests that the preconditioned Navier-Stokes equations exhibits hyperbolic-parabolic characteristics and hence applying reconstruction along a specific direction (local normal (LN) direction in our case) does not compromise on solution accuracy.

Another noteworthy feature of our newly proposed framework is the use of higher order flux discretization scheme which helps in accurately capturing the flow features at the solid-fluid interface. We use fifth-order WENO for this purpose. In recent times, it is emerging as an indispensable tool for modeling high speed compressible viscous flow involving shock obstacle interactions and high-speed moving body problems. Implementation of fifth-order WENO scheme demands that the flow field is extended to three layers of ghost cells inside the solid body through extrapolation. Chaudhiri et al.[7] and Pasquariello et al.[52] have utilized it for compressible inviscid flows. Recently Qu et al.[21] used it for simulating compressible viscous flows. We do not know of any other study other than these that have utilized such higher-order schemes to capture flow discontinuities.

This article is organized as follows. In section 2, the numerical methods employed in solving governing equations are presented. The details regarding immersed boundary pre-processing procedure are presented in section 3. A novel reconstruction strategy is proposed in section 4. Issues of spatio-temporal discontinuity in moving body problems and its relation to the issues of spurious oscillations and mass conservations are discussed in detail in section 5. Force calculations and rigid body dynamics of the coupled flows are presented in section 6. The performance of our novel reconstruction strategy through two case studies involving stationary and moving airfoil are presented in section 7. The validation studies are presented in section 8. Finally, the work has been summarized in section 9.

## *2. Numerical Method*

We use an in-house, parallel multi-block compressible flow finite volume code [62-72]. The unsteady, 3D Favre averaged N-S equations are solved on structured grids in generalized curvilinear coordinates. The inviscid fluxes are discretized using low diffusion flux splitting scheme [73]. Viscous fluxes are discretized using the standard second-order central differencing method. The solver achieves spatial accuracy using either second-order total variation diminishing (TVD) scheme or a fifth-order weighted essentially non-oscillatory (WENO) scheme [74]. A two-point Euler differencing is used to calculate pseudo-time derivatives and a three-point differencing scheme is utilized for the physical time advancement. The flow solver uses METIS to partition a general multi-block grid over the number of processors. Information among the processors communicated through MPI (Message Passing Interface). A very detailed description of the flow solver is provided in our recent work [72].

## 3. Immersed Boundary Pre-processing Procedure

*3.1 Immersed Geometry Description*

The immersed body is represented by unstructured triangular meshes. A common way to describe triangulations is by providing an element-vertex connectivity data structure as in standard file formats such as STL, Neu, etc. This information forms the basis for efficient signed distance calculations and surface tracking as well as interpolation algorithms. When dealing with moving body problems or fluid-structure interaction problems, the level of complexities involved in these geometric operations increases. To perform these operations in an efficient and robust manner, a comprehensive data structure is needed. We use half-edge data structure which is considered to be the most suitable data structure that provides all kinds of adjacent queries in constant time as well as provides the capability to edit the mesh when bad elements, gaps or cracks are encountered. We construct this data structure from the standard element-vertex connectivity using the approach proposed by Alumbaugh and Jiao [75].

*3.2 Node Classification*

Accurately classifying the Eulerian grid nodes (as solid, fluid, immersed boundary, and ghost nodes) is the most crucial step in the immersed boundary approach (**Fig.1**). In literature, one can find two different ways to classify the nodes. One using the signed distance function [3, 6, 35, 56] and another using ray-casting approach [39, 76]. For an immersed body which is closed, smooth and has orientable surfaces, the signed distance is calculated by finding the dot product between line projected from given point 'P' (see **Fig.2a**) onto the surface (at point 'P$_0$') and it's surface normal. Depending on the sign of the dot product, a given node can be classified as a solid or fluid node. But this process fails when the immersed boundary has sharp edges as in airfoil (**Fig.2a**). Let us consider point A from the shaded region. In order to classify the node, project a line from point 'A' to the surface. Notice that the line falls at the vertex of the surface where the surface normal is discontinuous. The dot product between surface normal $\hat{n}_2$ and the projected line are in the opposite direction; thus, the exterior fluid point will be wrongly marked as the interior solid point. The signed distance approach to classifying any point in the shaded region shown in **Fig.2a** will fail. It is worth to note that the triangular meshes are not C$^1$ continuous at its vertex and edges, thus the normal is undefined at those points. On the other hand, in the ray-casting approach (**Fig.1b-1c**), a ray is cast from a point (say F as in **Fig.1b**) and the number of intersections it makes with the surface is counted.

In this study, we use Moller and Trumbore ray/triangle intersection algorithm[77] which solves a linear system of equations to find the barycentric coordinates (u,v,w) (see **Fig.2c**) and the distance from the origin to the point of intersection 'P'. As long as the computed value fulfills the barycentric criteria, the intersection point is within the bounds of the triangle. Depending on whether the ray intersects the surface at odd or even number of times, the nodes are classified as exterior or interior. An axis-aligned bounding box (**Fig.2b**) is implemented to reduce the number of intersection tests as a large number of grid nodes is located outside it. All the nodes outside the bounding box are classified as fluid nodes.

While both signed distance approach and ray casting approach classify the nodes only as fluid or solid, a separate algorithm is required to tag the immersed boundary nodes and ghost nodes. These nodes are the nearest neighbor fluid nodes to the immersed surface on which the solution reconstruction is performed. Usually, a small search radius is defined with the centroid of the triangular element as an origin [6, 56]. Any nodes that fall within the search radius is tagged as immersed boundary nodes. We found that this approach leads to inconsistent tagging around sharp regions of geometry. In the present study, we perform a loop over all solid nodes, thereby checking the status of immediate neighboring nodes. If the immediate neighbor is fluid, then this node is tagged as immersed boundary nodes. Similarly, a loop over all the immersed boundary nodes is performed to identify its immediate solid neighbors. Those solid neighbors are tagged as ghost nodes, which will be used for field extension approach in case of moving body problem.

## 4. Novel Reconstruction Strategy for Sharp Interface Solution Reconstruction Schemes

Our novel reconstruction strategy is motivated by the need for a robust strategy for handling sharp features/edges of complex geometries. Before moving further, we would like to emphasize two crucial observations from the immersed boundary and computational geometry literature regarding the mathematical nature of the difficulty in handling the sharp features/edges of complex geometries.

1. David Eberly [78] suggests that for a given point, the closest distance to a triangle could be a vertex, an edge, or face itself. For example, Senocak et al. [37] when tested for Stanford Bunny geometry, found that 68.4% of the IB nodes were closest to triangular elements, 31.5% to the edges and 0.1% to the vertices. But except for triangle face, vertex or an edge has no well-defined normal as they are not $C^1$ continuous [57]. Thus signed distance function calculations fail to accurately classify the nodes that are near the region where normal is not well-defined.

2. A little-appreciated fact in the immersed boundary literature is the importance of the direction along which solution reconstruction procedures are usually performed. Traditionally the approach is to find the projection point of immersed boundary nodes/ghost nodes on the immersed surface and construct the reconstruction stencil parallel to surface normal or along the grid lines irrespective of the fact whether the projection point from the immersed boundary node is closer to the edge or vertex or the face. Thus when sharp-edged regions are encountered where they do not have a well-defined normal, reconstruction stencils have been applied in wrong directions leading to poor representation of those sharp regions. This ultimately results in the deterioration of solution accuracy.

By appreciating these facts, we propose a novel reconstruction strategy which involves:

1. Computing closest surface point to the given immersed boundary node and identifying whether the point is closer to the face, edge or vertex of the triangular element.
2. Finding the pseudo-normals for edge and vertex.
3. Apply reconstruction stencils in the direction parallel to the face normal or edge normal or vertex normal depending on whether a given point is closest to face, edge or vertex.

*4.1 Closest Surface Point Computation*

This is carried out as a two-step process. The first step involves finding a minimum bounding sphere for the triangular element (see **Fig.3a**) and storing its center and radius. This radius is then compared with the distance between the grid node and the center of this sphere. The one with the minimum difference is chosen. In the next step, we use David Eberly's 'Distance between point and triangle in 3D' algorithm [78] which defines a squared distance function (Q) for any point on the triangle,T to the point P.

$$Q(u,v,w) = |T(u,v,w) - P|^2 \tag{1}$$

This function is a quadratic in barycentric coordinates (u,v,w). The closest point is given by the global minimum of Q, which occurs when the gradient of Q equals zero. The challenge is to find whether this point is closest to the edge (R2, R3 and R4), vertex (R5, R6 and R7) or to the actual face (R1) itself (see **Fig.3(b)**). In all the three cases, finding the distance from a point to triangle translates into finding the distance to a line, a point or plane respectively. The index of the closest triangle face, edge, or vertex is stored along with the point of intersection with the help of half-edge data structure.

*4.2 Angle Weighted Pseudo-normals*

In this study, we define angle weighted pseudo-normal for vertices and edges of all the triangles based on the work of Andreas and Henrik [57]. **Fig.4 (a)** depicts the representation of angle weighted pseudo-normal of a vertex which is defined as

$$\hat{n}_v = \frac{\sum_i \alpha_i \hat{n}_i}{\left\| \sum_i \alpha_i \hat{n}_i \right\|} \quad (2)$$

Where 'i' denotes the number of incident faces and $\alpha_i$ is the incident angle.

In case of an edge (see **Fig.4 (b)**) between face 1 and 2, the angle weighted normal is defined as

$$\hat{n}_e = \pi \hat{n}_1 + \pi \hat{n}_2 \quad (3)$$

*4.3 Direction of the Reconstruction Stencil*

**Fig.5** illustrates the direction along which the solution reconstruction stencil is applied. The points P1, P2, and P3, are closest to with vertex V, edge e and face F of the triangular elements respectively. For reconstructing the solution at node P1, a line parallel to angle weighted vertex pseudo normal $\hat{n}_v$ projected onto the surface at P1′ is constructed. Similarly, for P2, a line parallel to edge normal $\hat{n}_e$ is projected onto the surface at P2′. For P3, a line parallel to the surface normal $\hat{n}_s$ is constructed.

*4.4 Boundary Conditions*

The precondition based flow solver solves for the primitive variables [72], $[p, u, v, w, T]^T$. Thus these flow field variables are to be reconstructed at immersed boundary nodes as well as at ghost nodes (in case of moving body problem) such that it satisfies the following boundary conditions.

Dirichlet boundary condition is given by the equation,

$$u_f(t) = \begin{cases} U_\Gamma(t) & \text{Moving body} \\ 0 & \text{Stationary body} \end{cases} \quad (4)$$

Where $u_f$ is fluid velocity vector, $U_\Gamma$ is the velocity of immersed body. This ensures no-slip boundary condition

Similarly, Neumann boundary condition is applied for ensuring non-penetration condition.

$$-n.\nabla p = -\left(\frac{\partial p}{\partial n}\right)_\Gamma \quad (5)$$

For cases involving subsonic flows

$$-\left(\frac{\partial p}{\partial n}\right)_\Gamma = \begin{cases} \vec{n}.\rho \dfrac{DU_\Gamma(t)}{Dt} & \text{Moving body} \\ 0 & \text{Stationary body} \end{cases} \quad (6)$$

For moving body cases involving Supersonic flow conditions, a normal force balance at the interface provides the pressure boundary condition,

$$-\left(\frac{\partial p}{\partial n}\right)_\Gamma = \frac{\rho_s u_{ft}^2}{R} - \rho_s a_n \quad (7)$$

where $u_{ft}$ is the tangential component of the velocity field at the immersed interface. $a_n$ is the normal component of acceleration of the interface $a_\Gamma$. Heat flux along the normal direction is assumed to be zero.

$$\left(\frac{\partial T}{\partial n}\right)_\Gamma = 0 \quad (8)$$

*4.5 Reconstruction Stencil for Immersed Boundary nodes in Incompressible Flows*

The incompressible flow fields are reconstructed using a quadratic interpolation stencil approach as described in [72] employed in the direction of local normal. Note that in our novel formulation, this normal could be surface normal or vertex normal or edge normal depending on where the projected point is located on the immersed surface. We use a Point-in-Cube algorithm akin to Point-in-Polyhedron ray casting algorithm described in **section 3.2** to strictly ensure that the selected neighboring points form a cubic control volume surrounding our point of interest. The order of accuracy of the interpolation scheme has been studied through grid convergence study and is verified to be of second order. The study is published here [70, 72].

*4.6 Reconstruction Stencil for Immersed Boundary nodes in Compressible Flows*

We have noticed that for compressible flows which involve discontinuities, such as shock waves, quadratic interpolations cause non-physical values at the IB nodes. Trying to fit a quadratic profile between them can lead to negative pressure or density which may lead to unphysical values at the interface. Hence, we use an Inverse Distance Weighted interpolation [72] for reconstructing IB nodes in Compressible flows.

## 5. Field Extension Strategy for Moving Body Problems

In order to address the issue of mass conservation and spurious oscillation in moving body problems, solid nodes that are immediate neighbor to the immersed surface are marked as ghost nodes and solutions from the outside field are extrapolated so that even when a grid node encounters abrupt change in its role and are exposed to the fluid there is a continuity maintained [13]. For incompressible flows, a quadratic extrapolation is used to extend the fields to ghost nodes. This is very similar to the methodology adopted in **section 4.5.** For compressible flows, a linear interpolation based on Image Point (**Fig.6**) approach is adopted [10, 72, 79]. This field extension strategy also helps in maintaining the order of accuracy of the flux schemes near the immersed interfaces. For TVD schemes, one layer of ghost cells is adequate as demonstrated in [72]. In the present study for a $5^{th}$ order WENO scheme, ghost cells are extended to three layers from the immersed interface (G1, G2 and G3 in **Fig.6**).

## 6. Force Calculations and Rigid Body Dynamics

The forces on the body are calculated by integrating the pressure $P$ and viscous stresses $\tau$ over the immersed surface $\Gamma$ using the expression

$$F_i = \int_{\Gamma} \left[ -P\delta_{ij} + \tau_{ij} \right] n_j d\Gamma \tag{9}$$

Where $F_i$ denotes force in $i^{th}$ cartesian co-ordinate. For obtaining information regarding pressure and shear stress on the immersed surface, an inverse distance weighted approach is used (procedure as explained in **section 4.5**).

In case of coupled flows (**Section 8.2.6**), where fluid interaction with immersed body results in rigid motions, the forces obtained through above procedure is utilized to update the information regarding velocity and position of the solid at the advanced time levels. Verlet integration approach is used for the purpose.

$$\begin{aligned}
U_{i,\Gamma}^{n+1} &= U_{i,\Gamma}^{n} + \frac{\Delta t}{M_{\Gamma}} F_i^n \\
x_{i,\Gamma}^{n+1} &= x_{i,\Gamma}^{n} + U_{i,\Gamma}^{n} \Delta t + \frac{(\Delta t)^2}{M_{\Gamma}} F_i^n
\end{aligned} \tag{10}$$

Where $U_i$ is the velocity vector and $x_i$ position vector in $i^{th}$ Cartesian coordinate direction.

## 7. Performance of Novel Reconstruction Strategy in handling sharp edges in complex geometries

Before presenting the validation studies, we would like to demonstrate the performance of our novel reconstruction approach based on HCIB/GC – RTLN algorithm (that is scheme based on combination of HCIB method and ghost cell method but based on ray tracing (RT) node classification and local normal (LN) information for choosing the direction for imposing reconstruction schemes) in handling sharp edges in complex geometries by comparing our results with the results based on standard HCIB/GC reconstruction algorithm (that uses signed distance approach for node classification and surface normal direction for reconstruction). The results are also compared against the experimental and other numerical studies available in the literature. Note that we have previously published works based on the standard HCIB/GC reconstruction algorithm [70-72]. Also, to handle the ambiguity arising in node classification procedure while performing signed distance calculation, a bounding box approach is used. The nodes that are in the wake outside this bounding box is marked as fluid nodes and no reconstruction procedure is performed on these nodes as is the standard practice [40, 80].

To assess the performance of our novel reconstruction strategy, we choose two cases : an uniform flow past a stationary NACA0012 airfoil at an angle of attack $\alpha = 10°$, Mach number $M_\infty = 0.8$ and Reynolds number Re = 500; an uniform flow past an NACA0012 airfoil pitching about its mid chord at a reduced frequency of *f\*=0.1* , Re = 3000 and $M_\infty = 0.006$. The mean incidence of the airfoil is $\bar{\alpha} = 30^o$, and the angular amplitude $\Delta\alpha$ is 15º. The instantaneous angle of attack is given by *α(t)=ᾱ-Δαcos(2πft)*. The computational domain is of size, $40C \times 30C$ with grid nodes $425 \times 317$ nodes.

In both the cases the sharp trailing edge is retained. No special trailing edge treatment is applied. For Flux reconstruction HCIB/GC – RTLN algorithm uses second-order TVD scheme. This is to ensure proper comparison with the results of standard HCIB/GC algorithm as it utilizes the second order TVD scheme.

The stationary case is in transonic flow regime, while the pitching airfoil problem is in ultra-low Mach number regime. We have chosen these two cases with extreme conditions to demonstrate the robustness of our immersed boundary frame work in handling sharp edged geometries in a wide variety of flow configuration. To the best of our knowledge, no other study has reported results either stationary or moving body based on the immersed boundary approach at such low Mach number of order $O(10^{-3})$. A more detailed discussion regarding the performance of solver at such low Mach numbers is presented in **section 8.**

*7.1 Stationary Airfoil with a sharp trailing edge:*

The results from our present HCIB/GC –RTLN algorithm and HCIB/GC algorithm are compared with the body-fitted results of Jawahar and Kamath [81] in **Fig.7.** Results from our present reconstruction framework provide very good match with [81] while that from HCIB/GC algorithm show that the streamlines do not strictly reattach itself to the airfoil surface at the trailing edge tip.

Moving further, we present the contour plots of vorticity, velocity and pressure contrasting the results obtained from HCIB/GC-RTLN and HCIB/GC algorithm (**Fig. 8**). As observed in **Fig.8c,** the solution from HCIB/GC algorithm deteriorates near trailing edge due to the presence of kink. Whereas, **Fig.8d** yields a smooth velocity distribution at the trailing edge tip. The pressure distribution pattern even in the leading edge is different for **Fig.8e** when compared with **Fig.8f.** Comparison of flow field results suggests that our present algorithm captures flow field at the sharp trailing edge more accurately and on par with the body-fitted grid results. Note that uniform grid spacing with the resolution of *C/285* (where *C* is chord) is maintained in the near body region. More details regarding pressure and skin friction distribution are presented in **section 8.1.2**.

## 7.2  Moving Body Problem Involving Airfoil with Sharp Trailing Edge:

In this section we demonstrate that HCIB/GC-RTLN algorithm that we have developed handles the sharp trailing edges more efficiently even in moving body problem retaining the sharp features at all-time instances. We also demonstrate that even in a coarse grid configuration (see **Fig.9 (a)**) the HCIB/GC-RTLN retains the solution accuracy at all the time instances by comparing it with the fine grid (see **Fig.9 (b)**) solutions from HCIB/GC algorithm. The coarse grid has mesh resolution of *C/250*, while the finer grid has a resolution of *C/500*. A comparative plot illustrating flow past the pitching airfoil at different time instances over a pitching cycle is presented in **Fig.10(a), (b)**. The solutions from the present HCIB/GC-RTLN algorithm is compared with the HCIB/GC algorithm, numerical study of Kumar et al. [82] and experimental study of Ohmi et al. [83]. The streamline plots depict the time evolution of vortex dynamics due to the airfoil pitching motion.

As the airfoil begins its up-stroke motion impulsively from its minimum incidence, the flow remains attached initially till $t^*=1$. After which the flow starts separating at the leading edge. This results in the formation of leading-edge vortex (LEV). Growth of trailing edge vortex (TEV) is also observed at $t^*=1.5$. Both LEV and TEV grows as the pitching motion progresses and around $t^*=2.5$ both starts coalescing into one big LEV by the end of the up-stroke ($t^*=2.5$). As the down-stroke begins, the growth of this vortex stops and gets detached from the airfoil surface ($t^*=3.5$). Down-stroke is characterized by formation of multiple vortices spanning the entire upper surface ($t^*=4.0$). By the time the airfoil reaches end of the down-stroke ($t^*=5.0$), the LEV is completely shed into the wake.

The challenge for any immersed boundary approach lies in accurately capturing the complex trailing edge vortex dynamics and the wake structures. This is possible only by retaining the sharp features of the trailing edge throughout the pitching cycle. For instance, consider the results of Kumar et al. [82] (the second column of **Fig.10**) obtained from the immersed boundary approach in a SOLA-MAC framework. Throughout the pitching cycle, it fails to reproduce any trailing edge vortex dynamics accurately. Throughout the up-stroke no TEV is observed ($t^*$=1.5-2.5). During the down-stroke a vortex is formed in the wake near the tip of the trailing edge ($t^*$=3.0). Consider results from the HCIB/GC algorithm (third column of **Fig.10**). This algorithm almost captures all the trailing edge vortex dynamics observed in the experiments [83]. But closer look at the time instances such as ($t^* = 2.5, 4.0$ and 5.0) suggests that there are extra vortices near the trailing edge tip which are not observed in the experiments. On the other hand, our present approach based on HCIB/GC-RTLN algorithm matches excellently with experimental observations at all the time instances.

While the streamline plots of both HCIB/GC and HCIB/GC-RTLN algorithms exhibit good agreement with the experimental data, a closer look at their wake structure provides better clarity regarding the nature of these algorithms in handling sharp edges. **Figs.11-15** report comparison of vorticity and pressure field between HCIB/GC algorithm and HCIB/GC-RTLN algorithm at three-time instances $t^*$ =0.5, 1.5 and 4.0. When an airfoil suddenly starts accelerating, i.e. pitching up, the trailing edge moves down, causing the counter-clockwise (positive) shear layer at the lower surface to roll up and shed into the wake. This vortex then convects downstream. Vorticity contour plots corresponding to $t^* =$ 0.5 (see **Fig.11**) reveals the presence of starting vortex in the downstream. The plots corresponding to HCIB/GC algorithm clearly reveals that the shear layer shed in the wake field is wavier and rolls up multiple times compared to the results from HCIB/GC-RTLN algorithm. A closer look into the trailing edge region suggests that the present approach reports a very smooth wake as expected.

The vorticity field corresponding to $t^*$=1.5 (see **Fig.12**) yields the positive shear layer corresponding to the results of HCIB/GC algorithm is highly unstable and keeps rolling up near the trailing edge tip and is shedding into the wake at regular intervals. Whereas the results corresponding to the HCIB/GC-algorithm reveals that the positive shear layer is stable and the starting vortex is continuing to convect downstream and decaying. A closer look into the trailing edge region depicts a discontinuity developed exactly at the trailing edge tip in case of HCIB/GC algorithm. However, our present approach (HCIB/GC –RTLN) exhibits a clean wake field. The pressure field for the same time instant is presented in **Fig.13.** HCIB/GC algorithm has a wavy pressure distribution as the flow reconstruction does not sharply impose the boundary conditions at the leading edge. Also, the flow field interface does not adhere to the real geometry even though the grid nodes are finer. Furthermore, one can see a kink in pressure value near the trailing edge tip. The results from HCIB/GC-RTLN on the other hand confirms a smooth pressure distribution, the sharp imposition of boundary conditions resulting in the flow field interface adhering to the real geometry. The trailing edge also shows smooth pressure field.

The vorticity field corresponding to $t^* = 4.0$ (see **Fig.14**) again confirms the presence of discontinuity near trailing edge in case of results corresponding to HCIB/GC algorithm. The shear layer that is rolling up to form trailing edge vortex is interrupted by a small region of negative vorticity near the tip of the airfoil. Also, the boundary layer on the upper surface which has negative vorticity is much stronger forming a vortex. The results from HCIB/GC-RTLN algorithm do not exhibit any such discontinuities and captures the smooth variation in the shear and boundary layers.

The pressure field (see **Fig.15**) does not reveal any discontinuity but one can notice the low-pressure region near the trailing edge is slightly shifted inwards in case of result corresponding to HCIB/GC algorithm. Also, near the trailing edge, the interface is thicker than the real geometry suggesting that boundary conditions are being not enforced sharply.

The results from HCIB/GC-RTLN on the other hand report smooth pressure field and sharp interface adhering to the real geometry.

The above discussion clearly demonstrates that our novel reconstruction algorithm (HCIB/GC-RTLN) is consistent in handling sharp edges both in static and moving conditions. Having reported the same, we move on to detailed validation cases in the following section which clearly establishes the robustness and capability of the present framework.

## 8. Results and Discussion

This section primarily has two subsections: The **section 8.1** deals with the validation study for three stationary cases. The **section 8.2** presents validation studies involving four moving body problems. In addition to the data presented in the manuscript, a supplementary data set is also provided for readers that contains more validation studies.

*8.1 Flow Past Stationary Objects*

*8.1.1 Impulsive Flow Past a Cylinder at Re = 3000*

This test case is presented to show that the present preconditioned solver with the IBM framework preserves time accuracy of the solutions at low Mach numbers. The impulsively started cylinder at Re = 3000 is simulated for inflow Mach number $M_\infty = 0.01$. At the inlet, uniform flow is specified while at the outlet a convective boundary condition is specified. A free stream condition is specified at the upper and lower boundaries. The computational domain is of the size $62D \times 46D$ with $1042 \times 857$ grid nodes. The near boundary regions are uniformly refined with the resolution of *D/320*.

The non-dimensional time is defined by the expression $T = 2U_\infty t/D$ as in [86]. **Fig.16** illustrates the comparison of the streamline pattern at three non-dimensional time instances ($T = 1.0, 3.0, 5.0$) with the experimental results of Loc et al. [84]. The pressure co-efficient distribution ($C_p = 2(P - P\infty)/\rho_\infty U_\infty^2$) corresponding to these time instances are also presented in **Fig.16.** The results show good agreement with the results of C.C.Chang et al. [85]. **Fig.17** depicts time evolution of vorticity at different non-dimensional time instances ($T = 1.4, 2.0, 4.0, 6.0$). The normalized surface vorticity distribution ($\omega *= \omega D/U_\infty$) corresponding to these instances are presented alongside. The normalized surface vorticity plots are in excellent agreement with P. Koumousatkos et al. [86].

*8.1.2 Viscous flow past Stationary Airfoil at High Subsonic and Transonic Regime*

The laminar flow past NACA0012 airfoil is simulated for two cases one at high subsonic and another at transonic flow regime: (1) $M_\infty = 0.5, Re_\infty = 5000$ at an angle of attack of $\alpha = 0°$, (2) $M_\infty = 0.8, Re_\infty = 500$ at an angle of attack of $\alpha = 10°$. The Reynolds number is based on chord $C$. The fluid domain for case-(1) is of size $40C \times 30C$ with the mesh size 895x378. The mesh spacing near the immersed boundary is *C/360*. **Fig.18 (a)** presents the streamlines near the trailing edge of the airfoil. The figure shows the presence of a pair of symmetric recirculation bubbles due to the flow separation near the trailing edge. This has been reported by Jawahar and Kamath [81]. **Fig.18 (b), (c)** exhibits pressure and skin co-efficient ($C_F = 2\tau_F/\rho_\infty U_\infty^2$) along the airfoil surface. The results match well with the works of Jawahar and Kamath [81]. Since the free stream parameters are near the upper end limit of the steady laminar flow, this is considered to be a difficult test case.

The fluid domain for case-(2) is of size $64C \times 51C$ with the mesh size $664 \times 556$. The mesh spacing near the immersed boundary is *C/285*. One can notice the presence of a large separated region at the upper surface of the airfoil (**Fig.18 (d)**) The vortex extends over 50% of the chord, and this is consistent with the observation of Jawahar and Kamath [81]. The quantitative comparison of the pressure and skin friction profiles (**Fig.18(e),(f)**) shows excellent match with the literature [81]. **Table 1** tabulates the drag and lift co-efficient comparing with the existing literature.

*8.1.6 Schardin's Problem: Moving Shock-Stationary wedge Interactions*

Here we consider a test case which involves the interaction of non-stationary $M_\infty = 1.3$ shock wave past a triangular prism. Schematic of the computational domain is presented in **Fig.19 (a).** The computational domain is of size $200mm \times 150mm$ with computational nodes of $1170 \times 850$. The flow field is initialized with moving shock using standard Rankine-Hugoniot relations [92]. The static pressure and temperature of the stagnant downstream field are taken to be 0.05MPa and 300K.

**Fig.19 (b)** illustrates the numerical schlieren image of waves arising out of the incident shock wave (IS) moving past the stationary wedge. The results correspond to time instant $t = 147.78\mu s$. The numerical results are in excellent match with the experimental observation [91] shown in the shadowgraph image **Fig.19(c)**. The unsteady evolution of these reflected shocks (accelerating, decelerating), expansion fans, multiple triple points, Mach stems, slip lines form the characteristic feature of this flow. Some instances of these evolutions can be observed from **Fig.20**.

**Fig.21** compares the trajectory of triple points (TP1 and TP2) and locus of the vortex center (V) obtained from the present simulation with that of Chang and Chang [91] and A. Chaudhuri et al. [7] . The results are in good agreement with the available studies.

**8.2 Moving Body Problems**

*8.2.1 Inline oscillating 2D Cylinder in Quiescent Medium*

To validate the present solver for a moving body problem at low speeds, an in-line oscillation of cylinder in a quiescent medium is simulated. The parameters are chosen from the experiments conducted by Dutsch et al. [89]. The motion is characterized by $Re = U_c D/\nu = 100$ and Keulegan-Carpenter number $KC = 2\pi X_0/D = 5$ where $D$ is the diameter of the cylinder, $\nu$ is the kinematic viscosity, $U_c$ is the maximum velocity with which cylinder oscillates, and $X_0$ is the amplitude of cylinder oscillation. The Mach number based on the $U_c$ is given by $M_c = 0.001$. The computational domain is of size $50D \times 30D$ with grid nodes $677 \times 432$ with grid resolution near boundary as D/100. The sinusoidal oscillation along the x-direction is given by the expression $X(t) = -X_0[sin(2\pi f_e t)]$, where $2\pi f_e t$ represents the phase position $\phi$, and $f_e$ is the frequency of the oscillating cylinder. **Fig.22** reports the velocity profile of the flow field in four locations ( $x = -0.6D, 0D, 0.6D, 1.2D$ ) and at three different phase positions $\phi = 180°, 210°, 330°$. The results are in excellent match with the experimental data [89]. The drag force variation (**Fig.23**) over a cycle compared with [89], the results are in an excellent match with the experimental and other numerical studies of [89].

*8.2.2 Steady Flow Past Transversely Oscillating Cylinder at low speeds.*

While the previous case has established the present solver has addressed the mass conservation and spurious oscillation issues, smooth pressure field, and time accurate wake field evolution, we present this test case to further examine the accuracy and robustness of the solver in handling more complicated flow pattern. This study is chosen following the work of Guilmineau et al. [93]. The objective is to reproduce the vortex switching phenomenon that occurs as the frequency of the oscillation changes and validate the time accurate evolution of near boundary pressure and surface vorticity. The flow is characterized by three important parameters: Reynolds number (based on the uniform velocity $U_\infty$ and Diameter $D$ of the cylinder) $Re = 185$, oscillation amplitude $Y_0 = 0.2D$, excitation frequency $f_e$. The free-stream Mach number for this flow is $M_\infty = 0.004$. The prescribed motion for the cylinder is $y(t) = Y_0 \cos(2\pi f_e t)$. Computational domain of size $100D \times 80D$ with $684 \times 556$ Cartesian grid nodes is adopted. Uniform grid spacing with the resolution of D/100 is maintained in the near body region.

The case is simulated for two different excitation frequencies $f_e = 0.8 f_0 \, \& \, 1.2 f_0$. $f_0$ is the vortex shedding frequency of the fixed cylinder. Guilmineau et al. [93] observed that for $f_e/f_0 < 1.0$, where the natural vortex shedding frequency dominates, the vortex shedding phenomenon happens periodically varying in a sinusoidal manner. For $f_e/f_0 > 1.0$, the length of the vorticity attached to the upper side of cylinder decreases, while that of the lower side increases.

**Fig. 24** reports flow past a transversely oscillating cylinder with two different frequencies $f_e = 0.8 f_0$ (left) and $f_e = 1.2 f_0$ (right). Vorticity contour, pressure distribution, and normalizes surface vorticity distributions for both the frequencies are presented alongside each other. By contrasting each other, one can notice that flow phenomena discussed in the above paragraph are reproduced in the present simulation. The pressure and normalized vorticity distributions too are in excellent agreement with the body fitted grid solutions of [93].The qualitative change in the flow field is also reflected in the time evolution of the lift and drag, as observed in **Fig. 24(d),(h)**. This pattern is also in good agreement with that of [93].

*8.2.3 Moving Cylinder in Compressible Viscous flow under supersonic conditions*

We now study the problem of moving cylinder at supersonic speeds in a viscous fluid. The study is designed as a Galilean invariant test of the stationary cases presented in supplementary data (**Section S.1**), i.e. the fluid remains stationary, and cylinder moves at constant supersonic speeds reaching a steady state. The case is simulated for two Mach numbers $M_D = 1.2$ & $2.0$ and Reynolds number $Re_D = 300$. Note that both the Mach number and Reynolds number are defined based on the cylinder velocity $U_D$. The computational domain is of the size $80D \times 16D$ with the grid nodes of $5740 \times 242$. The near boundary regions uniformly mesh, and refinement of *D/80* is maintained throughout the cylinder motion. The cylinder was initially placed at (1.5D, 8D). Initially, we present the numerical Schlieren image (**Fig.25**) of a cylinder moving at $M_D = 1.2$ & $2.0$ respectively showing the wake structure of the flow field behind the cylinder and development of bow shock ahead of the cylinder. As the cylinder covers the flow-through time, $t^* = tU_D/D = 60.0$ it reaches a steady state. The quantitative comparison of the pressure distribution along the moving cylinder (at $t^* = 60.0$) with that of the results obtained from stationary solutions as well as Takahashi et al. [94] exhibit excellent agreement. The difference between the stationary and moving body solutions are within the plotting distance of solution obtained by [94]. These results confirm that the present IBM framework works well for objects moving under high speeds incompressible viscous flows.

*8.2.6 Shock-driven Cylinder*

We now study a problem where a planar shock interacts with a rigid stationary cylinder driving it to move forward. This case was studied earlier by Henshaw et al. [95] using an overset grid technique and recently by Luo et al. [96] , Murualidharan and Menon [22]. The purpose of this test case is to demonstrate that solver's capability in handling coupling of the fluid and solid equations. The Schematic of the problem is presented in **Fig.26(a)**. The computational domain is of the size $0.4m \times 0.4m$ with $656 \times 656$ grid nodes. The cylinder with diameter $D = 0.1m$ is placed in the domain with its center at a location of $[x, y] = [0.15m, 0.2m]$. The grid spacing is of resolution *D*/164. The mass of the cylinder is 1.626e-6 Kg. A planar shock with Mach number M=1.5 relative to the flow ahead is initially located at $x = 0.5m$. The flow is at rest ahead of the shock with its initial conditions as

$$\rho_0 = 1.1825 kg/m^3, \gamma = 1.4, u_0 = 0.0, v = 0.0, P_0 = 101325 Pa$$

The state of the flow behind the shock is determined by Rankine-Hugoniot jump conditions [92] as,

$$\rho_1 = 1.56549 kg/m^3, u_1 = 241.103, v = 0.0, P_1 = 249090.287 Pa$$

**Fig.26(b)-(e),** depicts the snapshots of numerical Schlieren image of planar shock interacting with the stationary cylinder at various time instances $t^* = tU_S/D$ where the speed of shock $U_S = 520.783 m/s$. As the shock collides with the cylinder, the cylinder is set in forwarding motion and generates a reflected shock at the back of the cylinder ($t^* = 0.5$). The forward motion creates a compression wave ahead of the cylinder, which later develops into bow shock($t^*=1.0$). As the incident shock wave diffracts around the cylinder, it generates a Mach-stem, which in turn interacts with the reflected shock and bow shock resulting in two pairs of three intersections ($t^*=1.5$). The present numerical schemes accurately capture all the above complex and highly transient coupled flow phenomena.

**Fig.27(a)** presents the x-component of position (of the center of the cylinder), velocity and force as a function of time *t\**. The y-component of the force is close to zero, as reported in [95]. The symmetry of the flow field in **Fig.32(b)** also confirms this by reporting the density contours of the flow field at *t\*=1.0*. The results show very good agreement with the literature validating the capability of the present IBM framework in handling coupled flows.

## 9. Summary

We have presented a robust and efficient sharp interface immersed boundary framework that is capable of simulating arbitrarily complex geometries (static or in motion) in a viscous flow environment at arbitrary Mach number. The framework we have developed performs all the necessary geometry related calculations by adopting well established computational geometry algorithms. Our novel reconstruction strategy allows for robust handling of sharp edges without any need for resorting to complex treatments such as cut cell or adaptive mesh refinement. This is clearly demonstrated by comparing results (for case studies involving flow past stationary and moving airfoil) from our algorithm (HCIB/GC-RTLN) with a standard algorithm (HCIB/GC [72]) as well as other existing literature. We show that for a reconstruction based sharp interface approach, apart from robust node classification procedure and higher order boundary formulations, the direction in which the reconstruction is performed plays a major role in determining the solution accuracy, especially when sharp edged geometry is involved. Thus, our novel reconstruction approach can be very useful for modeling, for instance, realistic insect wings which have corrugation and sharp features. For improving mass conservation and suppressing spurious oscillations in case of moving body problems, we use ghost cell-based field extension treatment

    The framework is initially validated for stationary problems involving a wide range of flow conditions. Any compressible flow solver encounters loss of accuracy and convergence at very low Mach numbers. Our validation study involving, impulsively started flow past a stationary cylinder (at M=0.01 and Re = 3000) demonstrates that the immersed boundary framework we have developed for our preconditioned flow solver is capable of modeling both steady and unsteady flows at low speeds. The cases involving flow past a stationary airfoil case (at M=0.5, Re = 5000 and M =0.8, Re=500) reveals that the solver is capable of not only capturing flow physics at high subsonic and transonic regime but also handling sharp edged geometries. Under supersonic conditions case involving moving shock-wedge interaction demonstrates the solver is capable of modeling highly transient high-speed flow phenomena accurately. The results from all these cases match well the experimental and

numerical data available in the literature demonstrating the accuracy and effectiveness of the IBM framework developed.

Further, validation studies involving moving body problems are carried out. At low speed regime, an inline oscillating cylinder (M=0.001, Re = 100) and a transverse oscillating cylinder (M=0.004, Re =185) are presented. The results from these cases exhibit smooth pressure and vorticity distribution and suppression of spurious oscillations showing good agreement with the literature. To demonstrate the solver's capability to model high speed moving body problem, a case involving cylinders moving at constant supersonic speeds is presented as a Galilean invariant study. The near wake structure and pressure distribution match well with the stationary case establishing both the Galilean invariance as well as the accuracy of the solver. Finally, a coupled problem is simulated, which involves modeling of cylinder driven by shock impact. The results from these cases too, are found to be in excellent agreement with the numerical data available.

## 10. Conclusions

The present study demonstrates a robust and efficient sharp interface immersed boundary (IBM) framework, which is applicable for all-speed flow regimes and is capable of handling arbitrarily complex bodies (stationary or moving). The framework employs a combination of HCIB (Hybrid Cartesian Immersed boundary) method and GC(Ghost-cell) for solution reconstruction near immersed boundary interface to handle sharp edges of complex geometries. Further, the formulation involves a versatile interface tracking procedure based on ray tracing algorithm and a novel three step solution reconstruction procedure that computes pseudo-normals in the regions where the normal is not well-defined and reconstructs the flow field along those directions. The developed IBM framework is applied to a wide range of flow phenomena encompassing all-speed regimes (M=0.001 to M = 2.0).

A total of seven benchmark cases (three stationary and four moving bodies) are simulated involving various geometries (cylinder, airfoil, wedge) and the predictions are found to be in excellent agreement with the published results

**Acknowledgment**, The authors would like to acknowledge the IITK computer center (www.iitk.ac.in/cc) for providing the resources for performing the computation work, data analysis, and article preparation.

**Appendix A: Computational Time of the Present IBM Framework**

In order to demonstrate the computational efficiency of the present framework, A sphere immersed in a cube problem is assumed here. The sphere is made of 1200 triangular elements and cube made of Cartesian grid with $40^3$ nodes. Below Table:A1 shows computational time taken by the IBM preprocessing procedure (This includes node classification, tagging of closest surface points to the IB nodes, constructing interpolation stencils) for two different frameworks reported in this work: HCIB/GC [72] and the present framework HCIB/GC-RTLN . It shows that present framework is much more robust than the HCIB-SDSN framework. Note that all the simulations are performed on a system which is based on Intel Xeon E5-2670V 2.5 GHz 2 CPU-IvyBridge (20-cores per node) with 128 GB of RAM per node.

| No. of. Processors | HCIB/GC [72] | HCIB/GC-RTLN (present framework) |
|---|---|---|
| 1 | $1.6406 \times 10^{-1}$ (s) | $4.2960 \times 10^{-2}$ (s) |
| 2 | $6.6400 \times 10^{-2}$ (s) | $2.125 \times 10^{-2}$ (s) |
| 4 | $5.6638 \times 10^{-2}$ (s) | $1.0742 \times 10^{-2}$ (s) |
| 8 | $3.1250 \times 10^{-2}$ (s) | $5.3778 \times 10^{-3}$ (s) |
| 16 | $3.1250 \times 10^{-2}$ (s) | $5.3778 \times 10^{-3}$ (s) |
| 32 | $3.1250 \times 10^{-2}$ (s) | $5.3778 \times 10^{-3}$ (s) |

*Table A1: Comparison of computational time*

| Table 1: Comparison of lift and Drag co-efficients of Present study compared with the previous studies | | | |
|---|---|---|---|
| Case | Reference | $C_L$ | $C_D$ |
| M=0.5, Re = 5000, $\alpha = 0°$ | Jawahar and Kamath [81] | 0.05557 | 0.00 |
| | Venkatakrishnan [87] | 0.5568 | -- |
| | Crumpton et al.,[88] | 0.05610 | -- |
| | Qiu et al.[20] | 0.0610 | 0.00 |
| | Present Study | 0.05660 | 0.00 |
| M=0.8, Re = 500, $\alpha = 10°$ | Jawahar and Kamath [81] | 0.27216 | 0.49394 |
| | GAMM [89] | 0.243-0.2868 | 0.4145-0.517 |
| | Qiu et al, [20] | 0.2822 | 0.4323 |
| | Present Study | 0.2885 | 0.4966 |

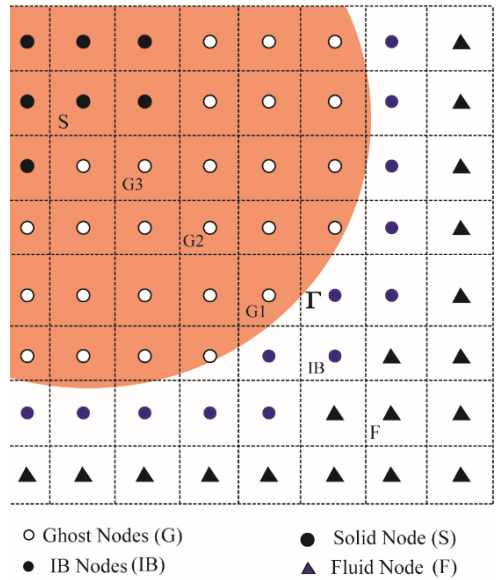

**Fig.1:** Schematic of Grid node classification with respect to the immersed body interface ($\Gamma$)

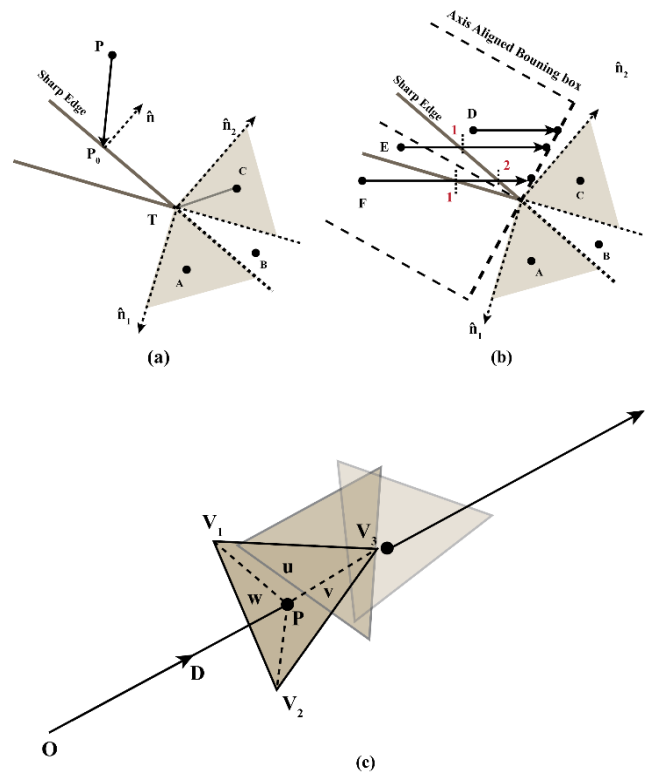

**Fig.2:** (a) Illustration of signed distance calculation for sharp edges (b) Illustration of ray-casting approach for node classification (c) Ray cast from origin O passes through a number of the triangle.

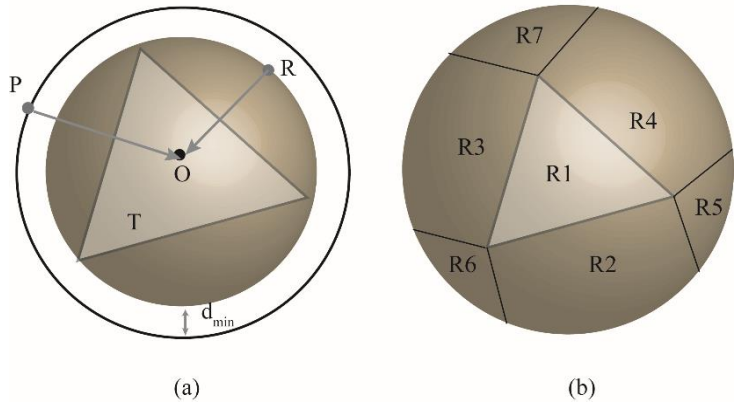

**Fig. 3**: (a) Triangle element (T) bounded by minimum bounding sphere with radius OR and a slightly larger sphere with radius OP. Here P denotes the nearest neighbor grid node. (b) Seven regions where the projected point from P could lie.

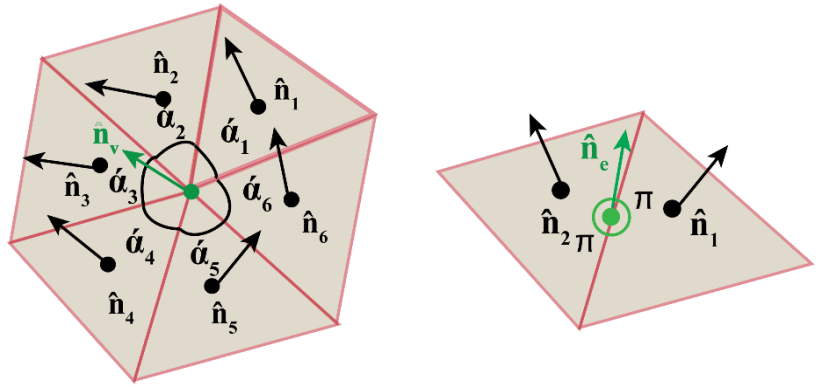

**(a)** **(b)**

**Fig. 4:** Representation of (a) Angle weighted Vertex normal (b) Angle weighted edge normal

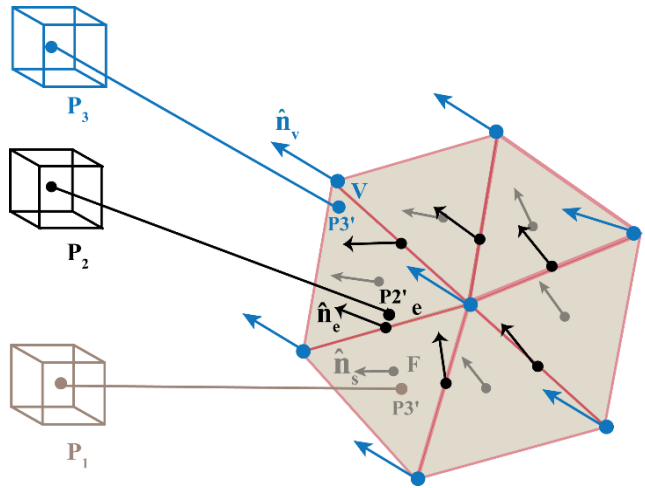

**Figure 5:** Representation of direction along which reconstruction stencil is applied.

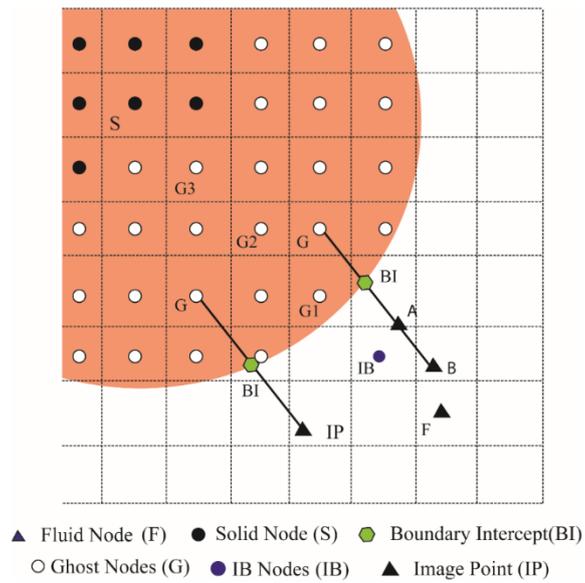

**Fig.6**: Field Extension Approach

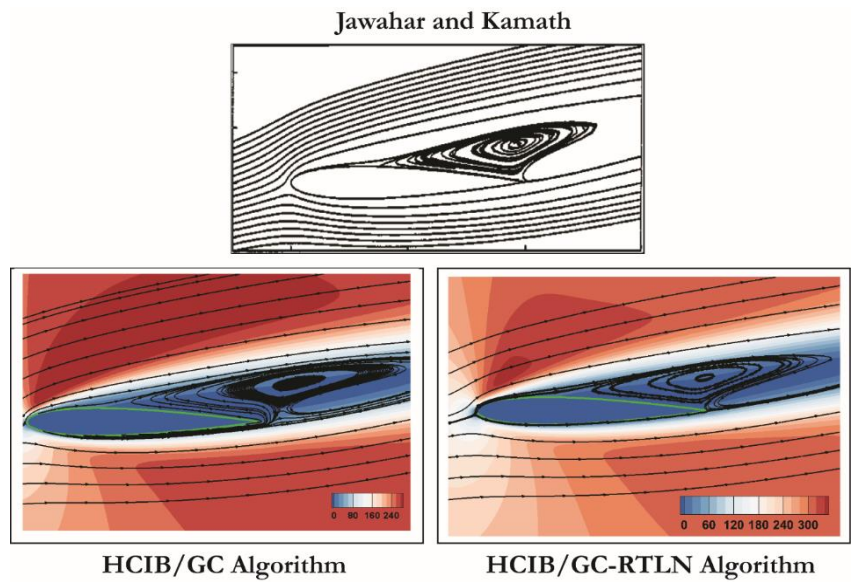

**Fig.7 :** Comparison of Streamline plot for transonic flow past a stationary airfoil. The results from HCIB/GC algorithm and HCIB/GC-RTLN algorithm is compared with body-fitted grid results of Jawahar and Kamath [81].

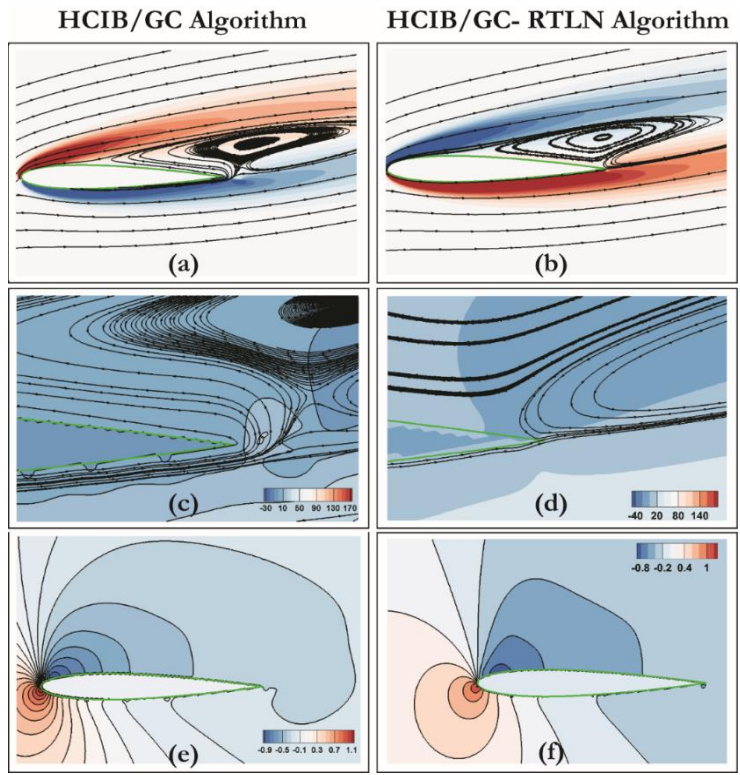

**Fig.8** : Comparison between HCIB/GC and HCIB/GC-RTLN algorithms : Vorticity contour (a-b); velocity along y-direction (c-d); pressure distribution (e-f)

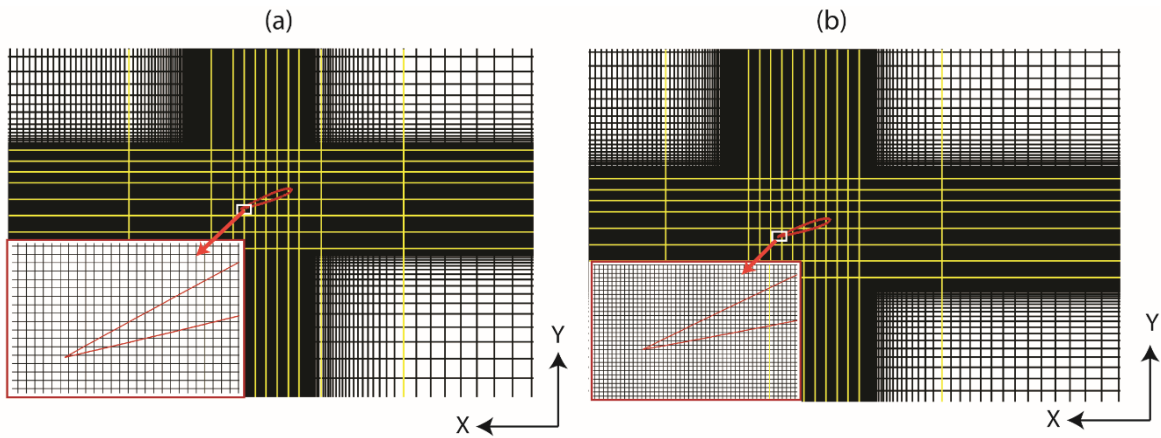

**Fig.9**: Grid Configuration for flow past pitching airfoil with a chord length of *C*: (a) Coarse grid with mesh resolution of *C/250*; (b) Fine grid with mesh resolution of *C/500*.

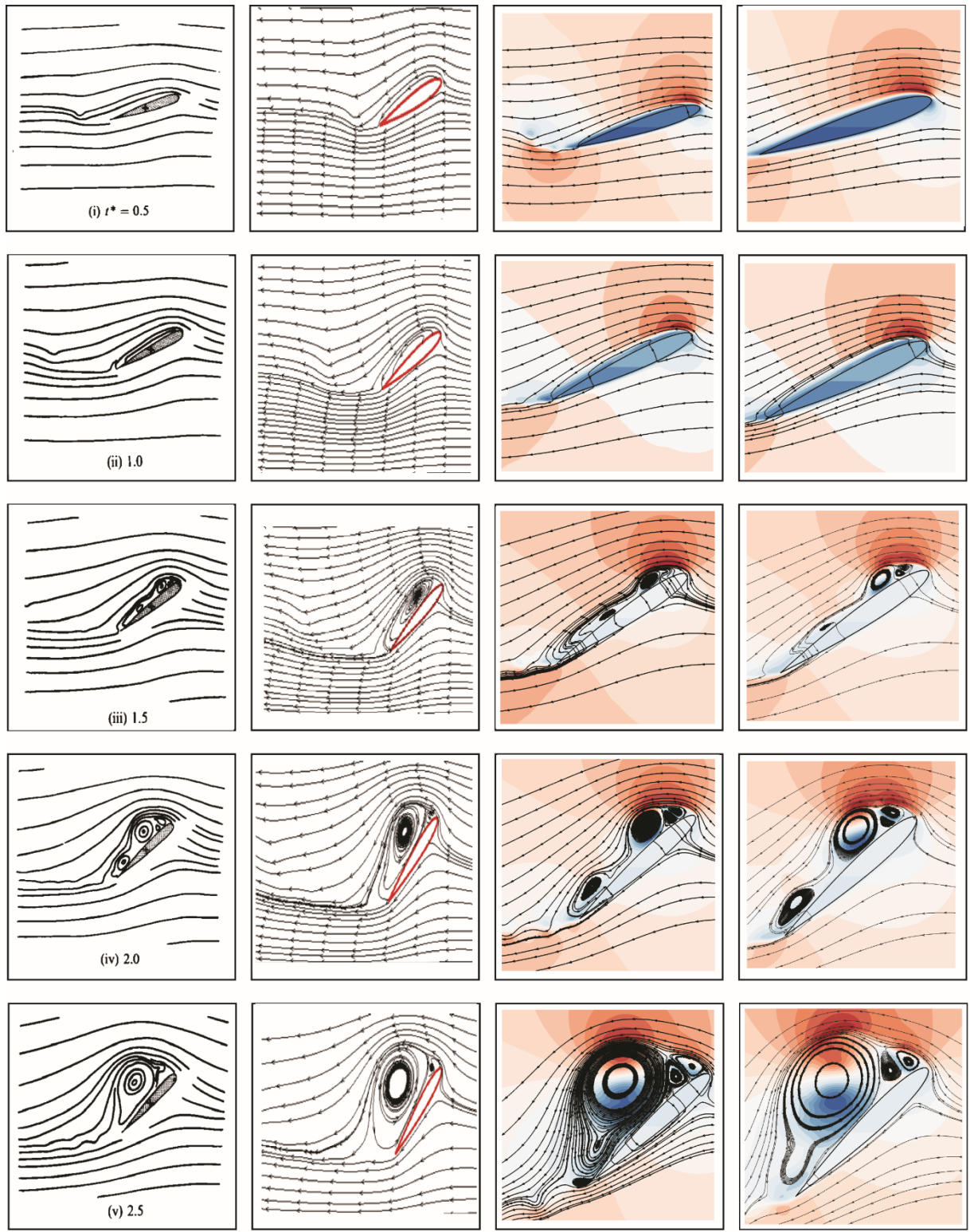

**Fig.10a** : Flow past pitching airfoil at $f^*=0.1$ , Re = 3000 and $M_\infty = 0.006$: The results from the present HCIB/GC-RTLN algorithm are compared with the solutions from HCIB/GC; Kumar et al. [82]; experimental results of Ohmi et al. [83]

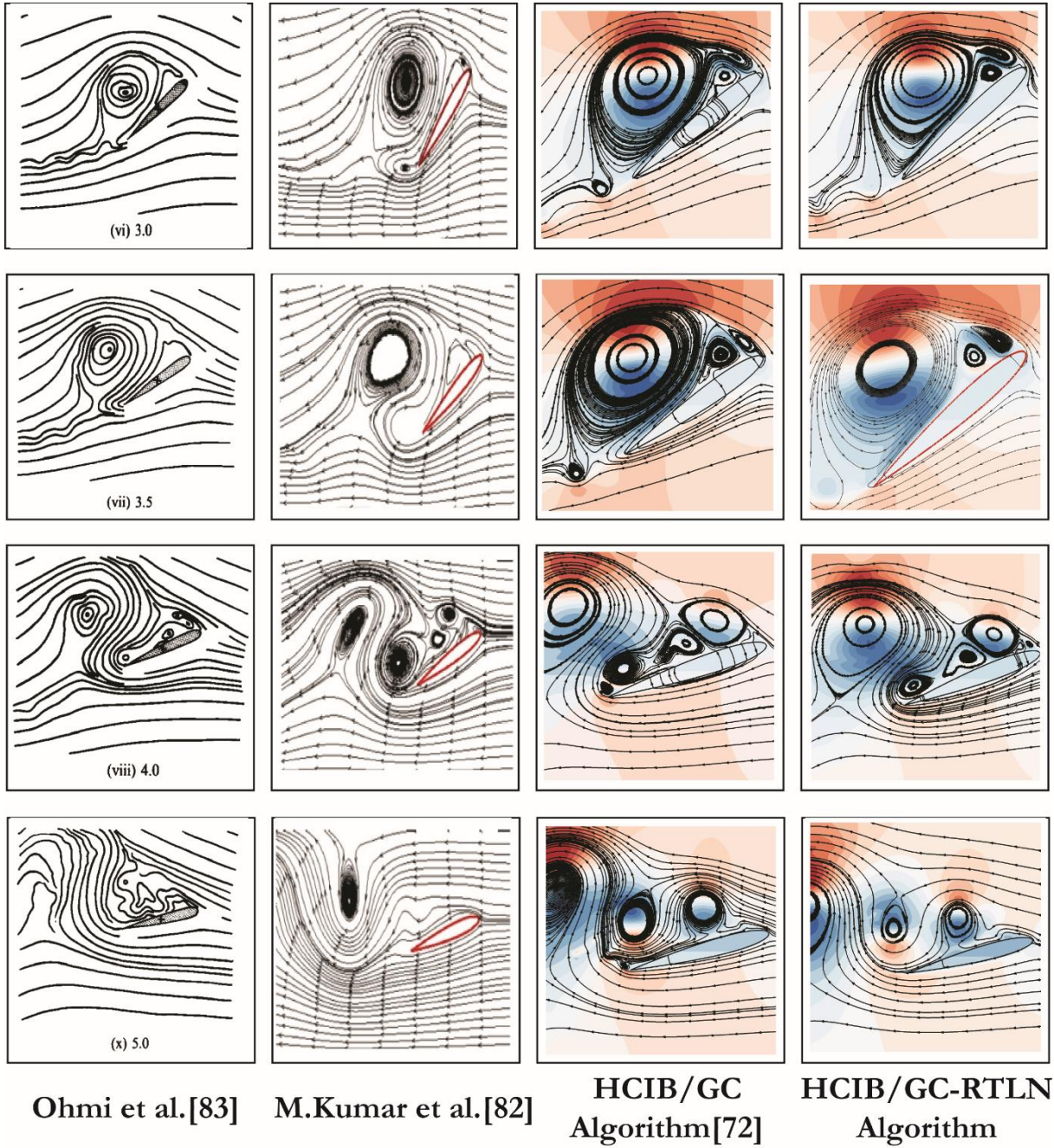

**Fig.10b** : Flow past pitching airfoil at $f^*=0.1$ , Re = 3000 and $M_\infty = 0.006$

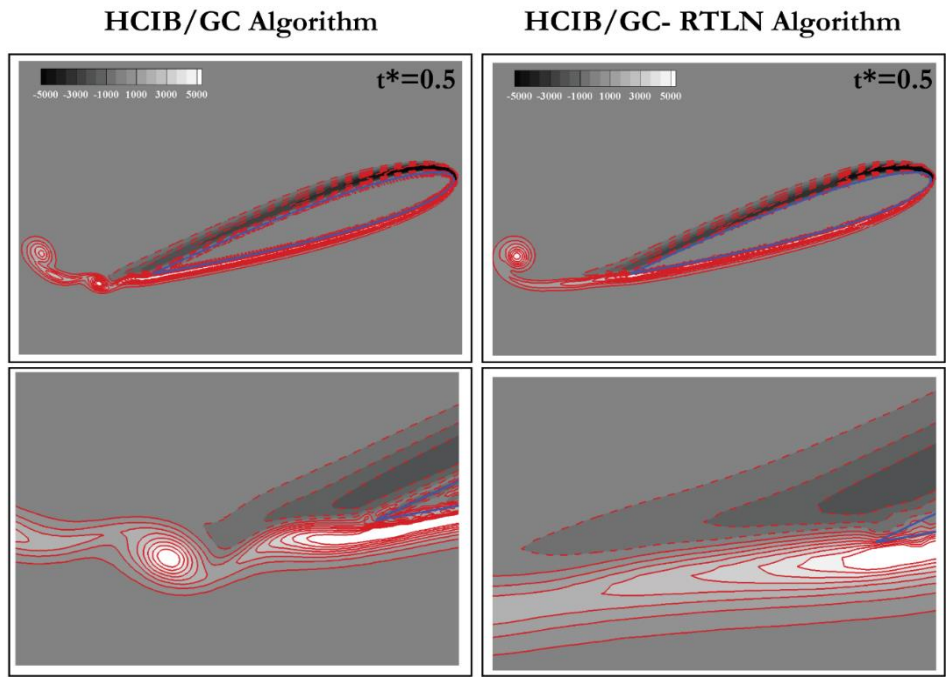

**Fig.11:** Comparison of HCIB/GC and HCIB/GC-RTLN algorithm at time instance $t^*$ =0.5. The contours correspond to the vorticity field.

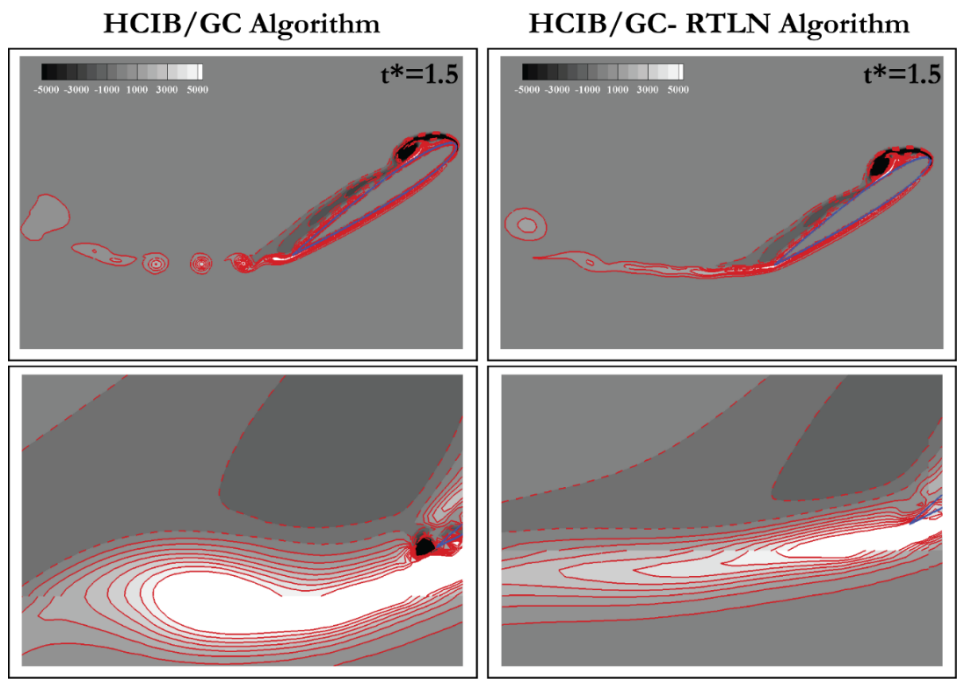

**Fig.12:** Comparison of HCIB/GC and HCIB/GC-RTLN algorithm at time instance t* =1.5. The contours correspond to the vorticity field.

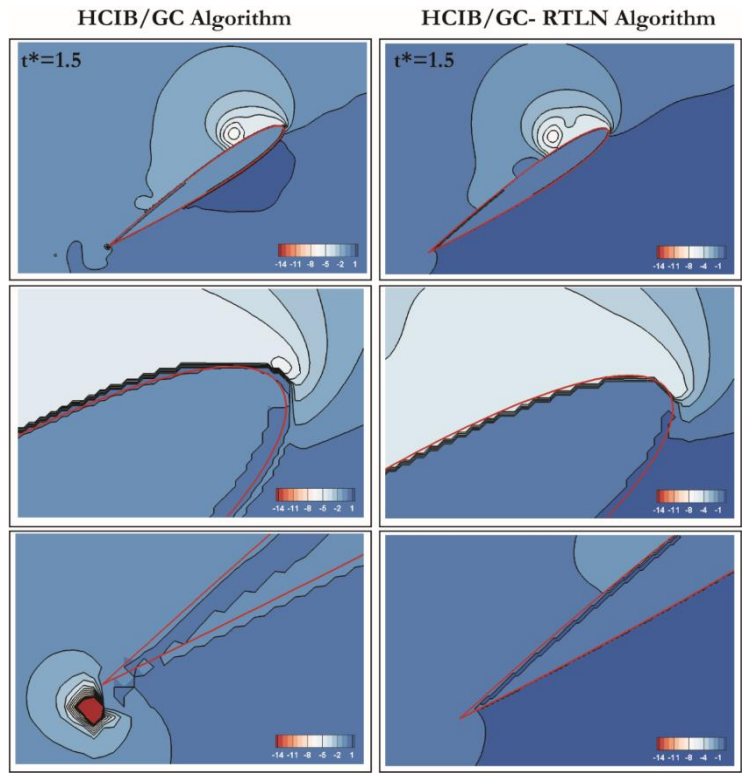

**Fig.13:** Comparison of HCIB/GC and HCIB/GC-RTLN algorithm at time instance t* =1.5. The contours correspond to the vorticity field.

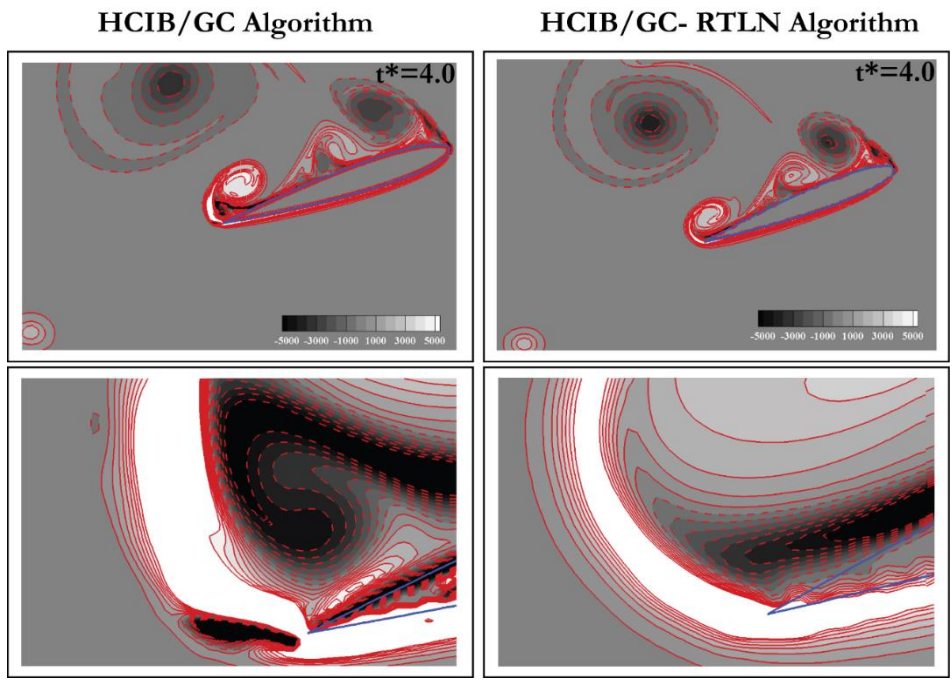

**Fig.14:** Comparison of HCIB/GC and HCIB/GC-RTLN algorithm at time instance t* =4.0. The contours correspond to the vorticity field.

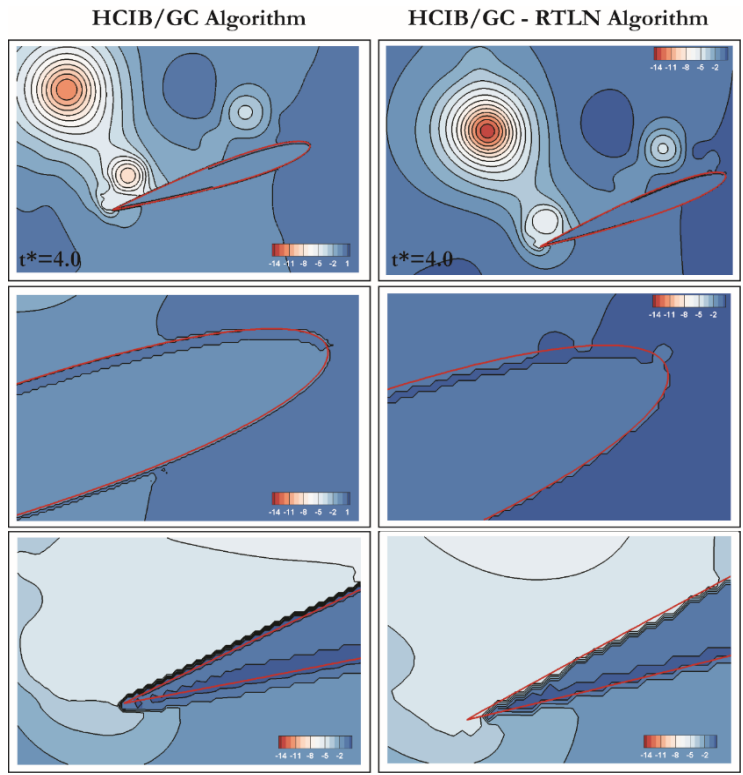

**Fig.15:** Comparison of HCIB/GC and HCIB/GC-RTLN algorithm at time instance t* =4.0. The contours correspond to the vorticity field.

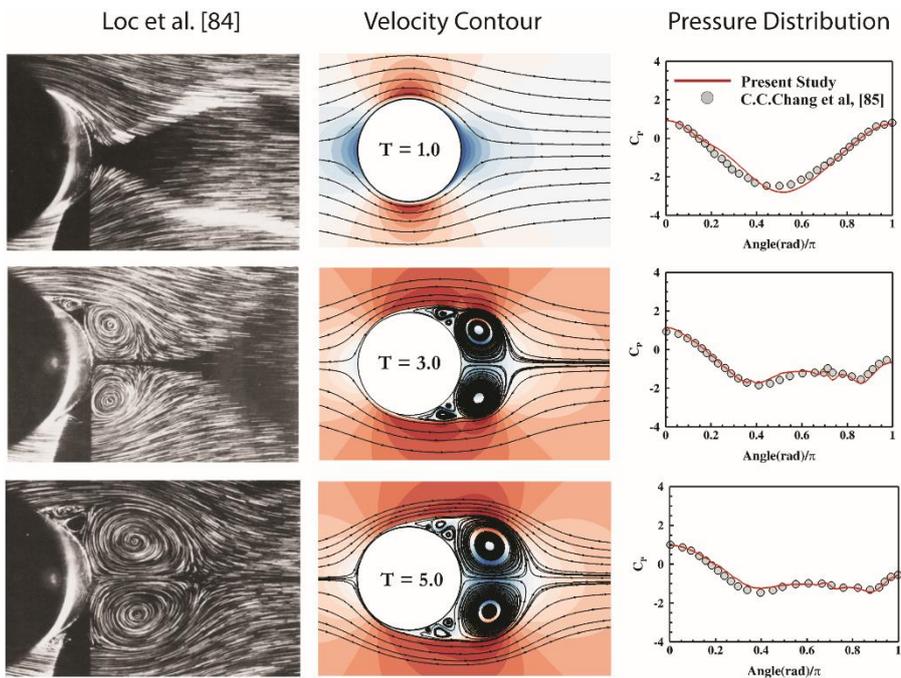

**Fig. 16** : Impulsively started flow past cylinder at Re = 3000. Experimental results of Loc et al. [84] are compared with the present results at three non-dimensional time instances ($T = 1.0, 3.0$ and $5.0$). Pressure distribution for corresponding time instances are compared with C.C.Chang et al.[85]

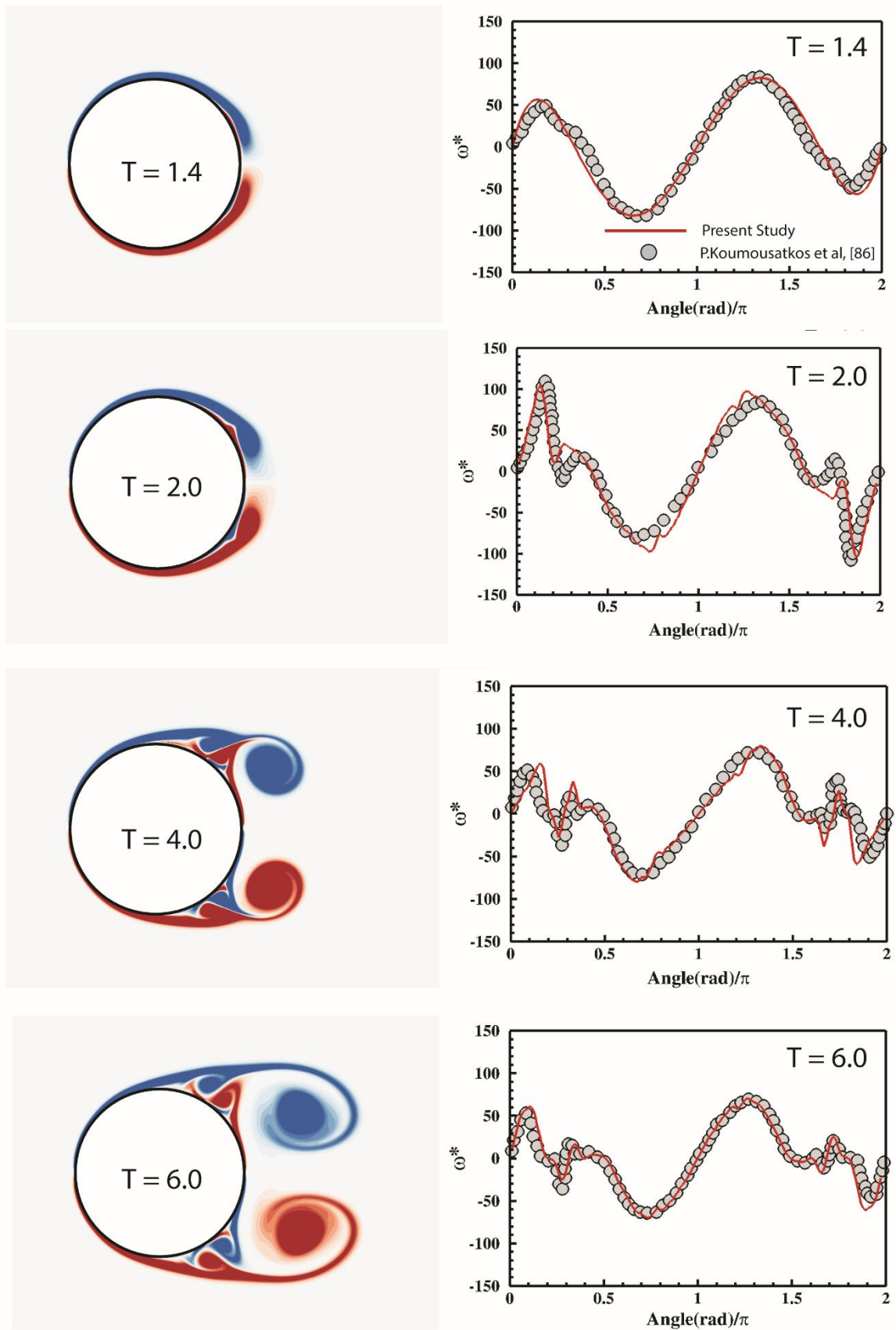

**Fig.17:** Instantaneous vorticity contour plots (left) for Re = 3000 at four instances T = 1.4, 2.0, 4.0, 6.0. The corresponding normalized surface vorticity (right) for these time instances are also presented. The plots are compared with the results of P. Koumousatkos et al. [86]

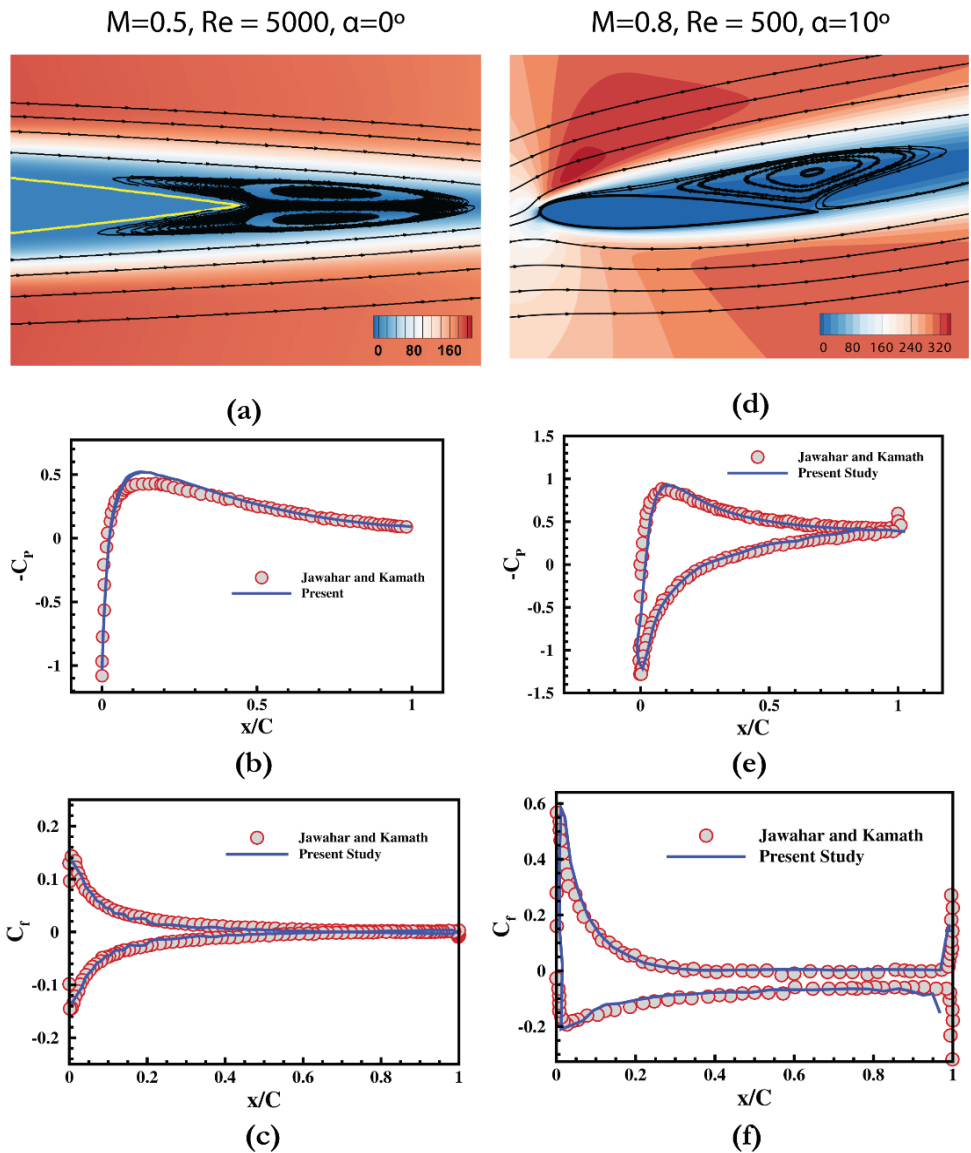

**Fig.18:** Viscous flow past NACA0012 airfoil at (1) M=0.5, Re = 5000, $\alpha = 0°$ (First Column) (2) $M_\infty = 0.8, \text{Re}_\infty = 500$, $\alpha = 10°$ (Second Column). Velocity contour plot with streamline distribtuions are shown in (a,d). Pressure co-efficient distribution along the surface (b,e). The pressure and skin friction distribution (c,f) along the surface are compared with Jawahar and Kamath [81]

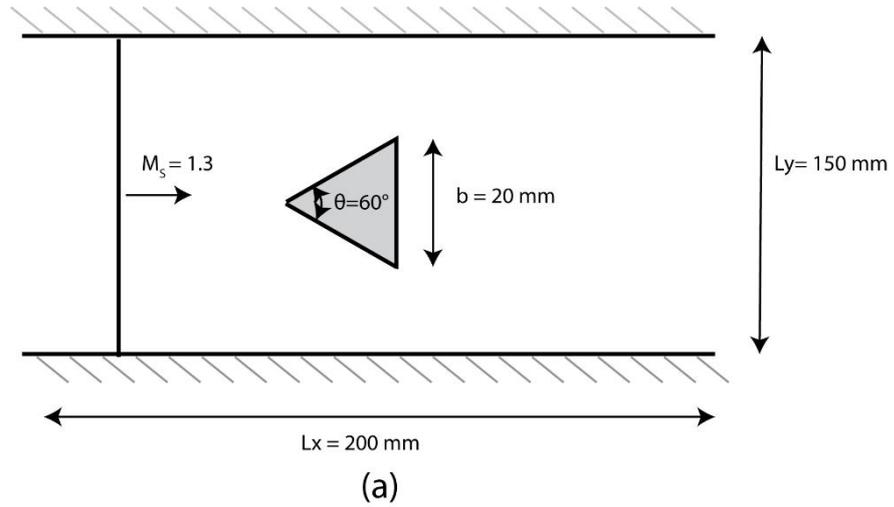

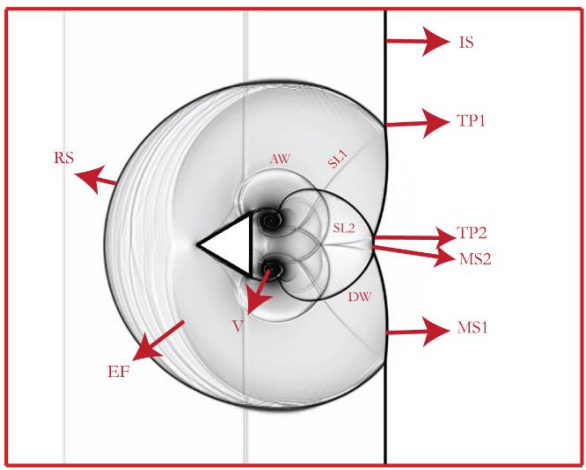

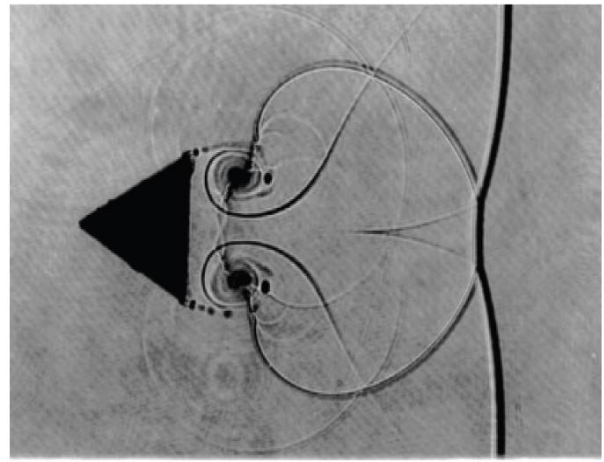

**Fig.19**: (a) Schematic of Shock-Wedge Interaction as proposed by Schardin [90] (b) Shadowgraph image of experimental result [91] showing various waves at $t = 151.0\mu s$ (c) Numerical Schlieren image of present result at $t = 147.78\mu s$ depicting different waves arising in Schardin's problem. RS : Reflected Shock; EF : Expansion Fan; IS : Incident Shock Wave; SL1,SL2: Slip lines; MS1,MS2- Mach Stems; TP1,TP2: Triple Points; AW : accelerated wave; DW- Decelerated wave; V-vortex

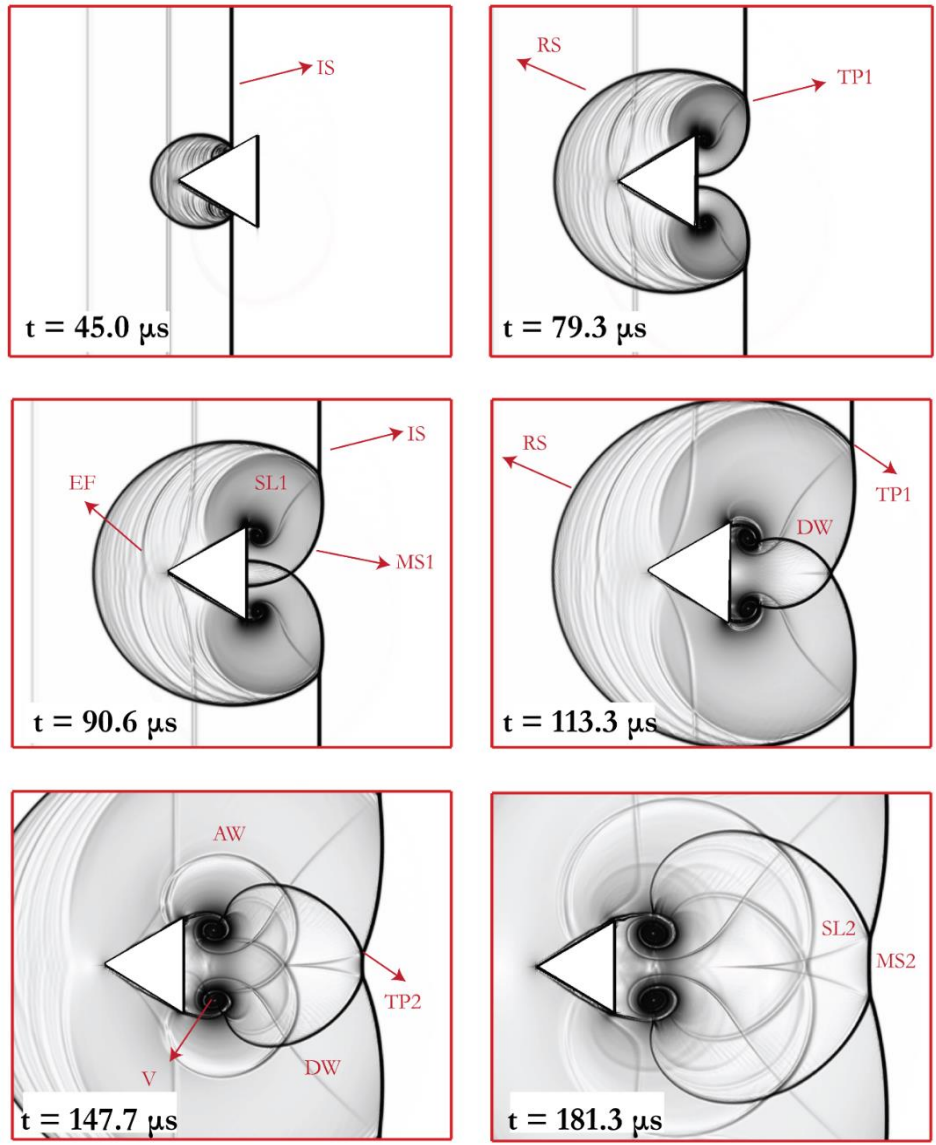

**Fig.20** : Snapshots of Numerical Schlieren Image for Schardin's problem at various time instances depicting unsteady evolution of various waves

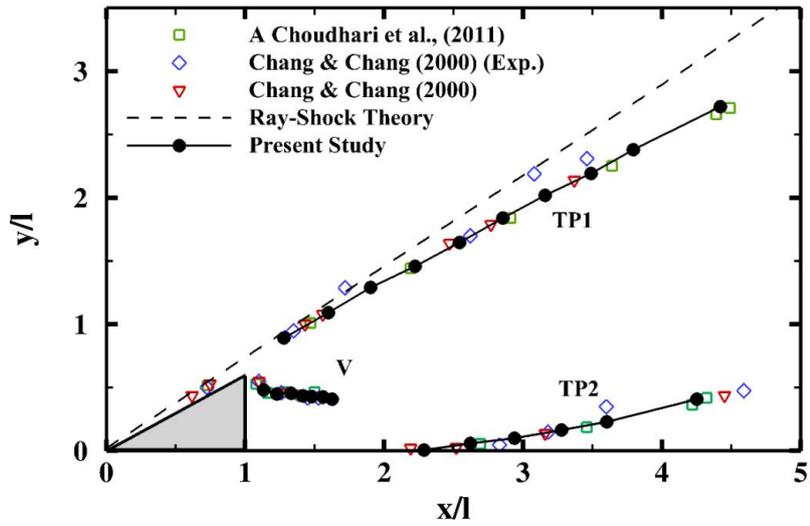

**Fig.21.** Comparison of the trajectory of triple points(TP1 and TP2) and locus of vortex center (V) with the literature [7, 91]

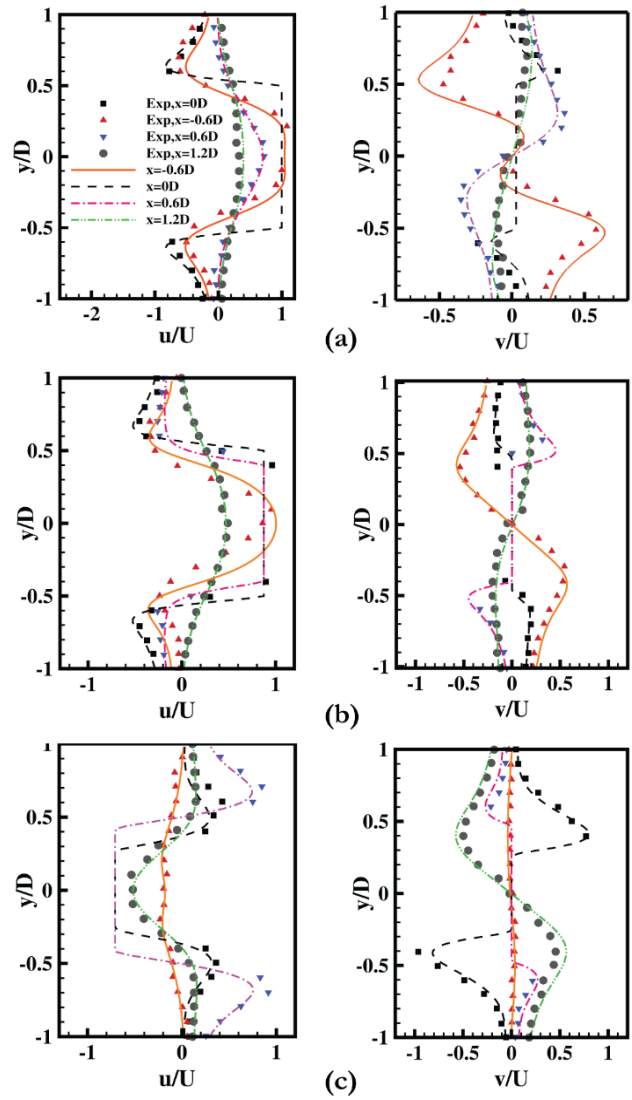

**Fig 22**: Comparison of Velocity components with experimental data of Dutsch et al. [89] at different phase positions a) $\varphi = 180°$ b) $\varphi = 210°$ c) $\varphi = 330°$. The symbols represent experimental data, and solid lines represent results from the present study.

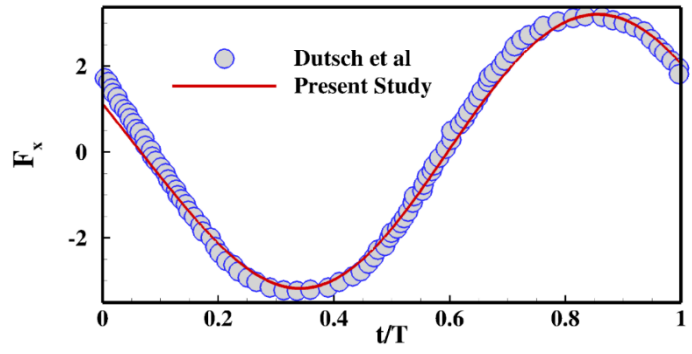

**Fig 23**: Comparison of Drag force over a cycle with experimental data of Dutsch et al. [89]

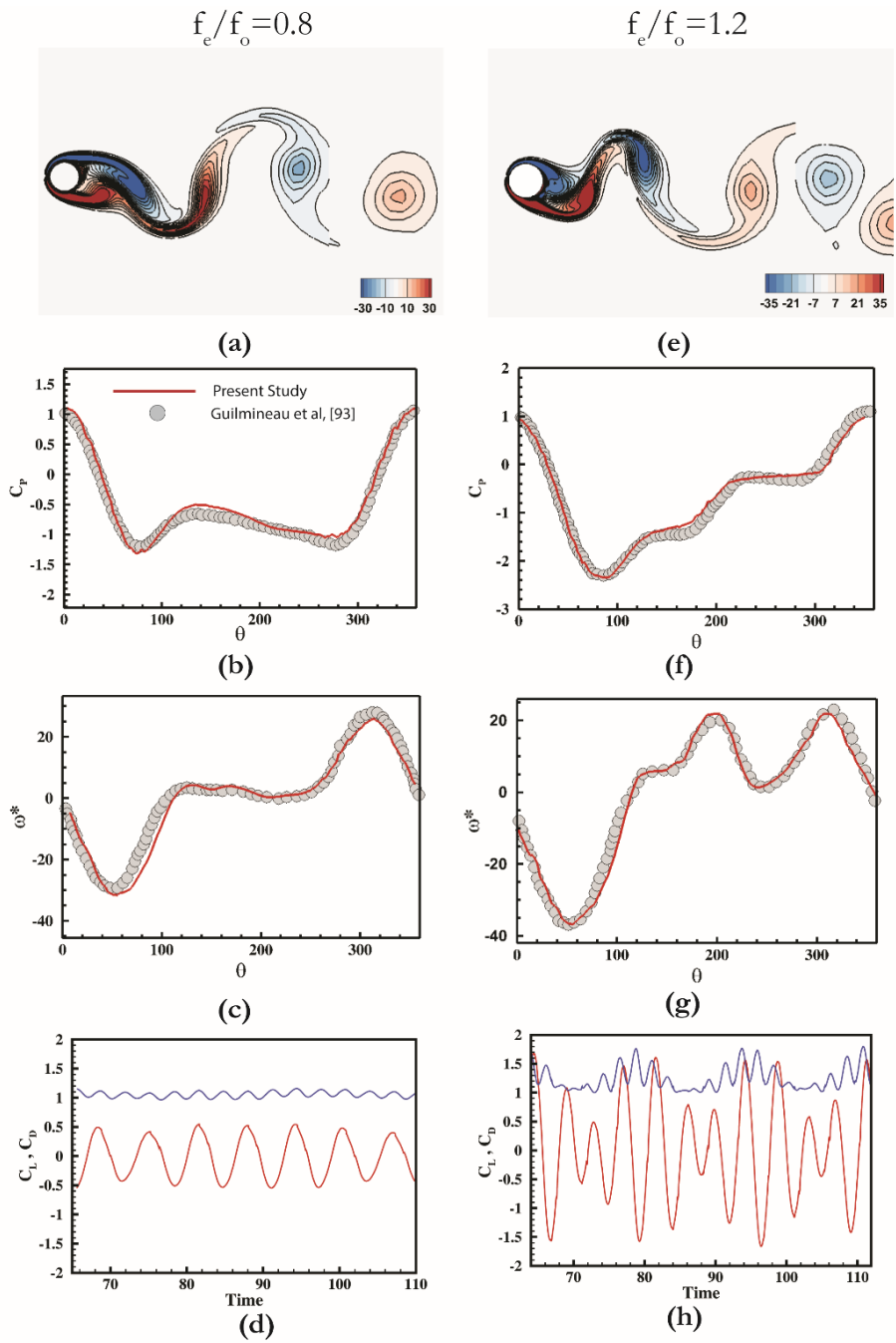

**Fig.24:** Flow past transversely oscillating cylinder for two frequency ratios $f_e/f_0 = 0.8$ (left) $f_e/f_0 = 1.2$ (right): (a), (e) vorticity plot ; (b),(f) Pressure distribution along the cylinder; (c),(g) normalized surface vorticity; (d),(h) Time history of lift and drag for transversely oscillating cylinder . The pressure coefficient and normalized surface vorticity are compared with the body fitted grid of [93]

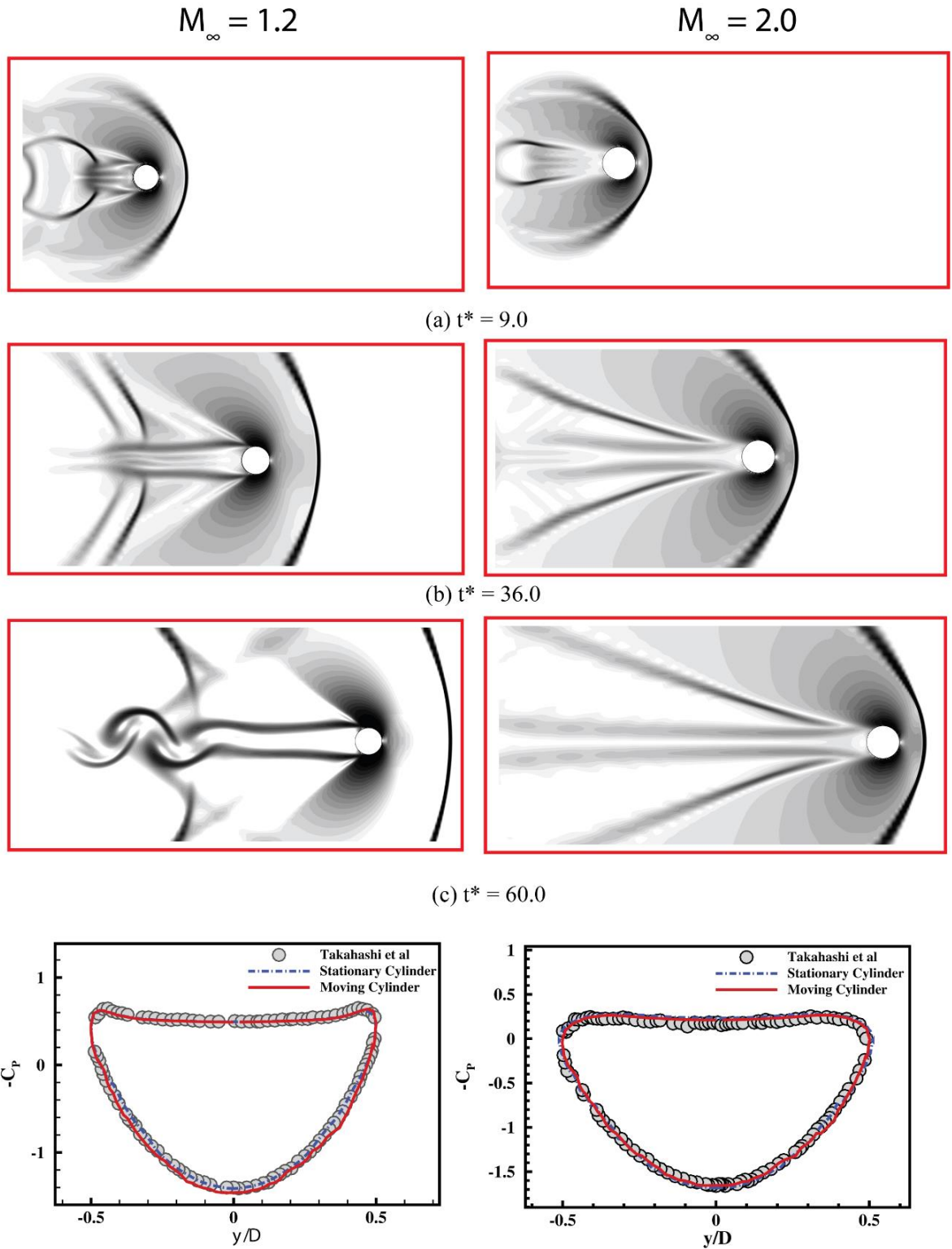

**Fig.25 :** Numerical Schlieren image of a cylinder moving at $M_D = 1.2$ and $2.0$ in a stationary viscous fluid at three time instants ($t^* = 9.0, 36.0, 60.0$). The last row depicts the pressure distribution along the cylinder surface at $t^* = 60.0$. The results are compared with Takashi et al.[94] and Stationary cylinder case simulated using present approach.

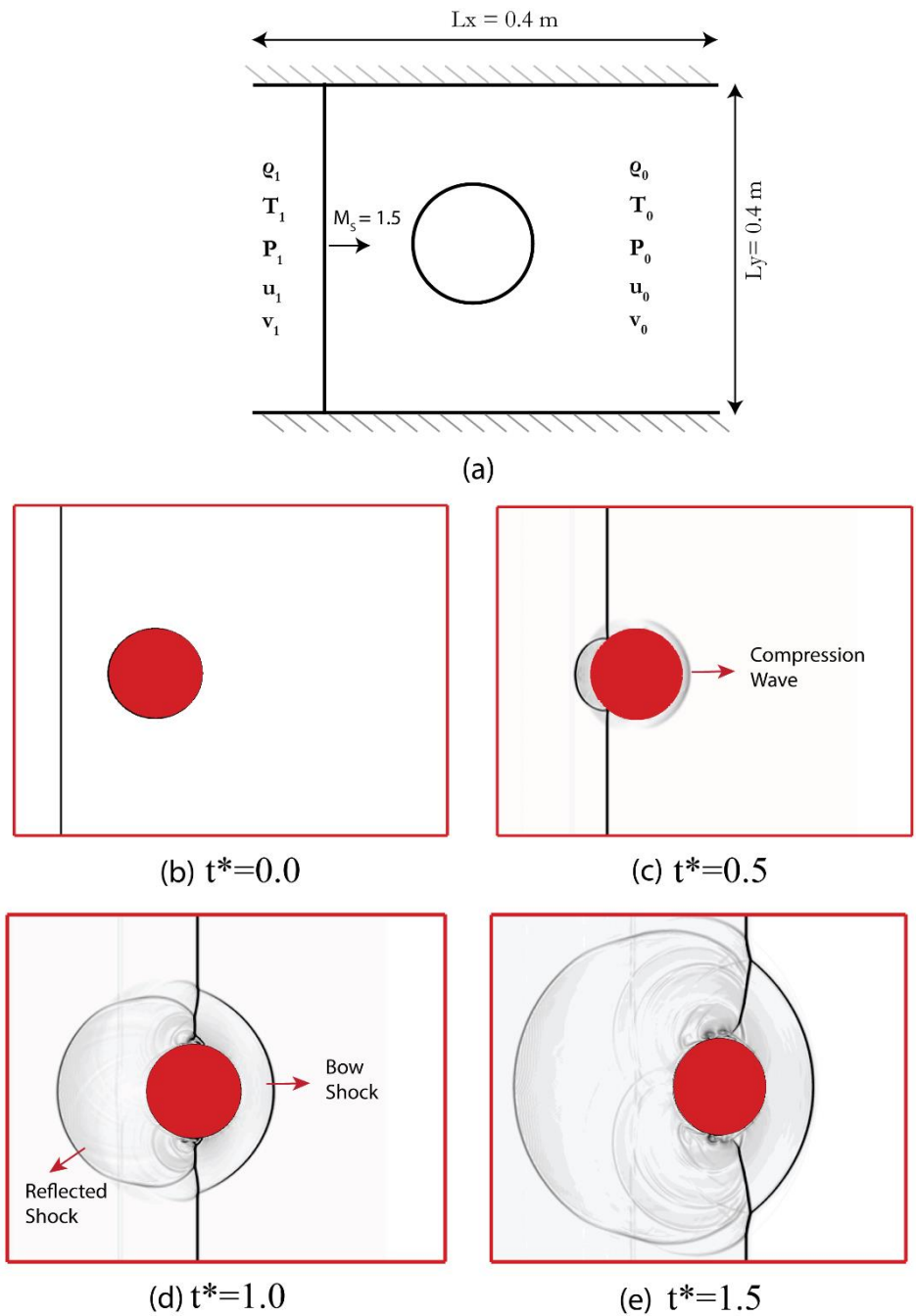

**Fig.26:** (a) Schemetic of shock driven cylinder case; (b)-(e) Numerical Schlieren Image of Shock driven cylinder at various time instances $t*$.

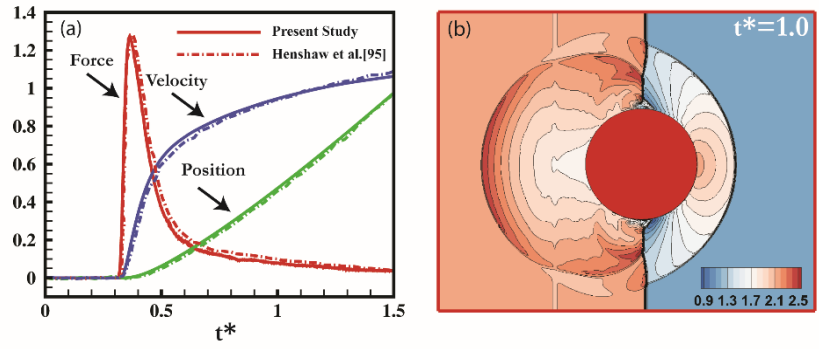

**Fig.27:** (a) Position, velocity, and force as a function of time compared with the result of Henshaw et al. [95] (b) Density field corresponding to t*=1.